\documentclass[aps,showpacs]{revtex4}
\usepackage{epsfig}
\usepackage{color}
\newcommand{\ben}{\begin{eqnarray}}
\newcommand{\een}{\end{eqnarray}}
\newcommand{\nnu}{\nonumber\\}
\newcommand{\bef}{\begin{figure}[htb]\centering}
\newcommand{\eef}{\end{figure}}
\newcommand{\bet}{\begin{table}[hbt]\centering}
\newcommand{\eet}{\end{table}}

\begin{document}
\title{Single transverse-spin asymmetry for $D$-meson production in semi-inclusive deep inelastic scattering}
\author{Zhong-Bo Kang and Jian-Wei Qiu}
\affiliation{Department of Physics and Astronomy, 
                 Iowa State University,  
                 Ames, IA 50011, USA}

\begin{abstract}
We study the single-transverse spin asymmetry for open charm production in the semi-inclusive lepton-hadron deep inelastic scattering. We calculate the asymmetry in terms of the QCD collinear factorization approach for $D$ mesons at high enough $P_{h\perp}$, and find that the asymmetry is proportional to the twist-three tri-gluon correlation function in the proton. With a simple model for the tri-gluon correlation function, we estimate the asymmetry for both COMPASS and eRHIC kinematics, and discuss the possibilities of extracting the tri-gluon correlation function in these experiments.
\end{abstract}                 
\date{\today}
\maketitle

\section{Introduction}
Single-transverse-spin asymmetries (SSAs) have received much attention in recent years, both experimentally and theoretically. Large SSAs have been consistently observed in various experiments at different collision energies 
\cite{SSA-fixed-tgt,SSA-dis,SSA-rhic}. As a consequence of the parity and time-reversal invariance of the strong interaction, the SSA is directly connected to the transverse motion of partons inside a polarized hadron.  Understanding the QCD dynamics behind the measured asymmetries should have the profound impact on our knowledge of strong interaction and hadron structure \cite{SSA-review}. 

Two QCD mechanisms for generating SSAs have been proposed \cite{Siv90, Efremov, qiu} and been applied \cite{qiu,Kanazawa:2000hz,Vogelsang:2005cs,siverscompare,Ans94,MulTanBoe,Boer:2003tx,Qiu:2007ar} extensively in phenomenological studies. Both of these mechanisms connect the SSA to the parton's transverse motion inside a transversely polarized hadron.  One mechanism relies on the transverse momentum dependent (TMD) factorization for the observed polarized cross sections, and explicitly expresses the SSA in terms of ``asymmetric" TMD parton distributions, known as the Sivers functions \cite{Siv90}.  The other follows the QCD collinear factorization approach for
cross sections when all observed momentum scales are much larger than the non-perturbative hadronic scale $1/{\rm fm}\sim \Lambda_{\rm QCD}$, and attributes the SSA to the twist-three transverse-spin dependent multi-parton correlation functions \cite{Efremov, qiu}.  Unlike the first mechanism, in which the SSA measures the TMD parton distribution and the spin-dependence of parton's transverse motion at a given momentum, the twist-three transverse-spin-dependent correlation functions reveal the net spin-dependence of parton's transverse motion when its transverse momentum is integrated.  Naturally, the moment 
of the spin-dependent TMD parton distributions could be related to the twist-three multi-parton correlation functions \cite{Boer:2003cm}.  Although two mechanisms each have their own kinematic domain of validity, they describe the same physics in the region where they overlap \cite{UnifySSA}.

Most studies in both mechanisms have been concentrated on the SSAs generated by either the quark Sivers function \cite{Vogelsang:2005cs,siverscompare,Ans94,MulTanBoe,Boer:2003tx} or the twist-three quark-gluon correlation function \cite{qiu,Kanazawa:2000hz,Qiu:2007ar,Kouvaris:2006zy}, which is defined as
\ben
T_F(x, x)=\int\frac{dy_1^- dy_2^-}{4\pi}e^{ixP^+y_1^-}
\langle P,s_\perp|\bar{\psi}(0)\gamma^+\left[ \epsilon^{s_\perp\sigma n\bar{n}}F_\sigma^{~ +}(y_2^-)\right] \psi(y_1^-)|P,s_\perp\rangle \, , 
\een
with the gauge links suppressed.  Possibilities of accessing the transverse motion of gluons, or, the gluon Sivers functions have also been investigated recently \cite{Anselmino:2004nk, Boer:2003tx}. Likewise, the spin-dependence of the gluon's transverse motion in the QCD collinear factorization approach is represented by the twist-three tri-gluon correlation function, $T_G(x,x)$, defined as
\ben
T_G(x, x)=\int\frac{dy_1^- dy_2^-}{2\pi}e^{ixP^+y_1^-}\frac{1}{xP^+}\langle P,s_\perp|F^+_{~~\alpha}(0)\left[ \epsilon^{s_\perp\sigma n\bar{n}}F_\sigma^{~ +}(y_2^-)\right] F^{\alpha+}(y_1^-)|P,s_\perp\rangle \, .
\label{TG_correlation}
\een
Its contribution to the SSA was first studied by Ji in the context of direct photon production in hadronic collision \cite{Ji:1992eu}.  Although direct-photon production provides a nice possibility to access $T_G(x,x)$, the extraction of $T_G(x,x)$ could be difficult due to the contribution from the quark-initiated subprocesses, and the limited knowledge on $T_F(x,x)$.  In this paper, we study the single-transverse spin asymmetry for open charm production in the semi-inclusive deep inelastic scattering (SIDIS) and argue that the SSA in SIDIS is a clean observable to extract the twist-three transverse-spin dependent tri-gluon correlation function, $T_G(x,x)$.

The $D$-meson production at large enough transverse momentum $P_{h\perp}$ in SIDIS is dominated by the photon-gluon fusion subprocess, $\gamma^*+g\to c+\bar{c}$, since the intrinsic charm contribution \cite{Brodsky:ic} is less relevant at large $P_{h\perp}$ and the photon-charm subprocess, $\gamma^*+c\to c+g$, is suppressed by the small charm quark distribution at the collision energy that we are interested in this paper.  We calculate the SSA for $D$-meson production in lepton-proton SIDIS in terms of the QCD collinear factorization approach, and find that the asymmetry is directly proportional to the diagonal part of the twist-three tri-gluon correlation function in the polarized proton, $T_G(x,x)$, because the photon-gluon fusion subprocess at this order does not have the so-called ``hard-pole'' contribution to the asymmetry \cite{UnifySSA}.  Therefore, the measurement of SSA for $D$-meson production in SIDIS is a direct measurement of the tri-gluon correlation function, $T_G(x,x)$. With a simple model for the tri-gluon correlation function, obtained under a similar assumption guiding the modeling of the quark-gluon correlation function, $T_F(x,x)$, we find that the asymmetry for both COMPASS \cite{compass} and eRHIC \cite{Deshpande:2005wd} kinematics is sizable and could be measured experimentally.  Recently, COMPASS experiment successfully measured the gluon polarization, $\Delta G$, in a longitudinally polarized proton based on the photon-gluon fusion process by tagging charmed mesons \cite{compass}. This certainly makes the measurement of SSA of open charm production in the same experimental setting and the extraction of the transverse-spin dependent tri-gluon correlation function, $T_G(x,x)$, promising.

We find that the SSA of $D$-meson production in SIDIS has a minimum at $z_h\sim 0.5$, and it increases as $z_h$ is moving away from this central value.  This increase of the SSA away from $z_h\sim 0.5$ has the same physics origin as the observed increase of the magnitude of SSA in hadronic pion production as a function of increasing $x_F$ (or rapidity $y$), and is a prediction of the twist-three formalism of the QCD collinear factorization approach to the SSA.

We also find that the twist-three gluonic contribution to the SSA has both similarity and difference from the twist-three fermionic contributions.  Both gluonic and fermionic twist-three contributions to the SSA have the so-called ``non-derivative'' and ``derivative'' terms, which correspond to the terms that are proportional to the twist-three correlation functions and the derivative of the correlation functions, respectively.  As noticed in Refs.~\cite{Kouvaris:2006zy,Koike:proof}, the fermionic ``non-derivative'' and ``derivative'' terms can be combined together and the dependence on the non-perturbative twist-three correlation function, $T_F(x,x)$, is proportional to a simple combination, $T_F(x,x)-xT_F'(x,x)$.  However, our explicit calculation shows that the ``non-derivative'' and ``derivative'' gluonic contribution can not be combined into the same simple form due to the difference in the partonic hard parts of these two terms. 

The same approach discussed in this paper can be applied for studying the SSA in open charm production in hadronic collisions, which is dominated by the gluon-gluon fusion if $T_G(x,x)$ is not too small comparing to the $T_F(x,x)$ \cite{Yuan:2008it,Vitev:2006bi,Kang:hadron}.  With the extraction of the tri-gluon correlation function, $T_G(x,x)$, and the knowledge of $T_F(x,x)$, we enter a new era of exploring non-perturbative physics beyond the parton distribution functions (PDFs) which have been well-studied in the past thirty years.  
 
The rest of our paper is organized as follows.  In Sec. \ref{ssa calculation}, we present our calculation of the SSAs for open charm production in SIDIS. We first introduce the relevant kinematics of open charm production in SIDIS and present the formula for the unpolarized cross section.  We then derive the twist-three formula for the SSA in QCD collinear factorization approach and express the asymmetry in terms of the tri-gluon transverse-spin dependent correlation function, $T_G(x,x)$.  We close this section by a discussion on the calculation of the color factor of the partonic hard part which depends on the color contraction of three gluon fields in the definition of the tri-gluon correlation function.  In principle, there could be two gauge invariant tri-gluon correlation functions, $T_G(x,x)$ and $\widetilde{T}_G(x,x)$ defined later, due to two independent ways to neutralize the color of the three gluon fields in the matrix elements.  We point out that only one of them, $T_G(x,x)$, could be related to the gluon Sivers function \cite{Anselmino:2004nk}.  In Sec. \ref{numerical}, we estimate the production rate of open charm mesons in SIDIS for both COMPASS and eRHIC kinematics.  We choose a simple ansatz for the tri-gluon correlation function, $T_G(x,x)$, and present our predictions for the SSAs of open charm production at the existing COMPASS experiment and the planned eRHIC experiment.  Finally, we conclude our paper in Sec. \ref{conclusion}.

\section{Calculation of single-spin asymmetry}
\label{ssa calculation}

We start this section by specifying our notation and kinematics of SIDIS. We consider the scattering processes of an unpolarized lepton, $e$, on a polarized hadron, $p$,
\ben
e(\ell)+p(P, s_\perp)\to e(\ell')+h(P_h)+X, 
\een
where $s_\perp$ is the transverse spin vector defined below, $h$ represents the observed $D$ meson with momentum $P_h$ and mass $m_h$.  For the collision energy that we are interested in this paper, we work in the approximation of one-photon exchange, and define the virtual photon momentum $q=\ell-\ell'$ and its invariant mass $Q^2=-q^2$.  We adopt the usual SIDIS variables:
\ben
S_{ep}=(P+\ell)^2, \qquad 
x_B=\frac{Q^2}{2P\cdot q},\qquad
y=\frac{P\cdot q}{P\cdot \ell}=\frac{Q^2}{x_B S_{ep}},\qquad 
z_h=\frac{P\cdot P_h}{P\cdot q}.
\een
It is also convenient to introduce the ``transverse'' component of the virtual photon momentum, $q$, as
\ben
q_t^\mu=q^\mu-\frac{q\cdot P_h}{P\cdot P_h}P^\mu-\frac{q\cdot P}{P\cdot P_h}P_h^\mu,
\een
which is a space-like vector and orthogonal to both $P$ and $P_h$. We define
\ben
\vec{q}^{\,2}_\perp\equiv -q_t^\mu q_{t\mu}=Q^2\left[1+\frac{1}{x_B}\frac{q\cdot P}{P\cdot P_h}\right]-\frac{m_h^2}{z_h^2}.
\label{qT}
\een
To completely specify the kinematics, we will work in the so-called {\it hadron frame} \cite{sidis}, where the virtual photon and the polarized proton are taken to have only one spatial component that is in the $z$-direction:
\ben
P^{\mu}=P^+\bar{n}^{\mu},  \quad\quad  q^{\mu}=-x_B P^+ \bar{n}^{\mu} +\frac{Q^2}{2x_B P^+}n^{\mu},
\een
where the light-cone momentum component is defined as $P^{\pm}=(P^0\pm P^3)/\sqrt{2}$, and $\bar{n}^{\mu}=(1^+,0^-,0_\perp)$, $n^{\mu}=(0^+,1^-,0_\perp)$ are two light-like vectors with $\bar{n}\cdot n=1$. The momentum of final-state $D$-meson can be written as
\ben
P_h^\mu=\frac{x_B P^+}{z_h Q^2}m_{h\perp}^2 \bar{n}^{\mu}+\frac{z_h Q^2}{2x_B P^+}n^\mu+P_{h\perp}^\mu,
\een
where $m_{h\perp}^2=m_h^2+P_{h\perp}^2$ with $P_{h\perp}=\sqrt{\vec{P}_{h\perp}^2}$. From Eq. (\ref{qT}) one can show that $q_\perp\equiv \sqrt{\vec{q}_\perp^{\,2}}=P_{h\perp}/z_h$ in this hadron frame, independent of mass $m_h$.

In this hadron frame, usually, one chooses the coordinate system such that the virtual photon has a vanishing energy component, corresponding to $P^+=Q/\sqrt{2}x_B$, and $P_{h}$ lies in the $xz$-plane (known as the {\it hadron plane}), as shown in Fig.~\ref{frame}. The lepton momenta, $\ell$ and $\ell'$ define the {\it lepton plane} and can be expressed in terms of variables $\psi$ and $\phi$ as follows \cite{sidis},
\ben
\ell^\mu&=&\frac{Q}{2}\left(\cosh\psi, \sinh\psi \cos\phi, \sinh\psi \sin\phi, -1\right),\nnu
\ell'^\mu&=&\frac{Q}{2}\left(\cosh\psi, \sinh\psi \cos\phi, \sinh\psi \sin\phi, +1\right),
\een
where $\phi$ is the azimuthal angle between the hadron and lepton plane, as indicated in Fig.~\ref{frame}, and 
\ben
\cosh\psi=\frac{2x_B S_{ep}}{Q^2}-1=\frac{2}{y}-1.
\een
We parametrize the transverse spin vector of the initial proton $s_\perp$ as
\ben
s_\perp=(0,\cos\phi_s, \sin\phi_s,0),
\een
where $\phi_s$ is the azimuthal angle of $s_\perp$ measured from the hadron plane, as shown in Fig.~\ref{frame}. If one uses the lepton plane as the reference to define the azimuthal angle of $s_\perp$ as $\Phi_S$, and that of hadron plane as $\Phi_h$, one has the relation $\phi_s=\Phi_S-\Phi_h$ and $\phi=-\Phi_h$.
\bef
\psfig{file=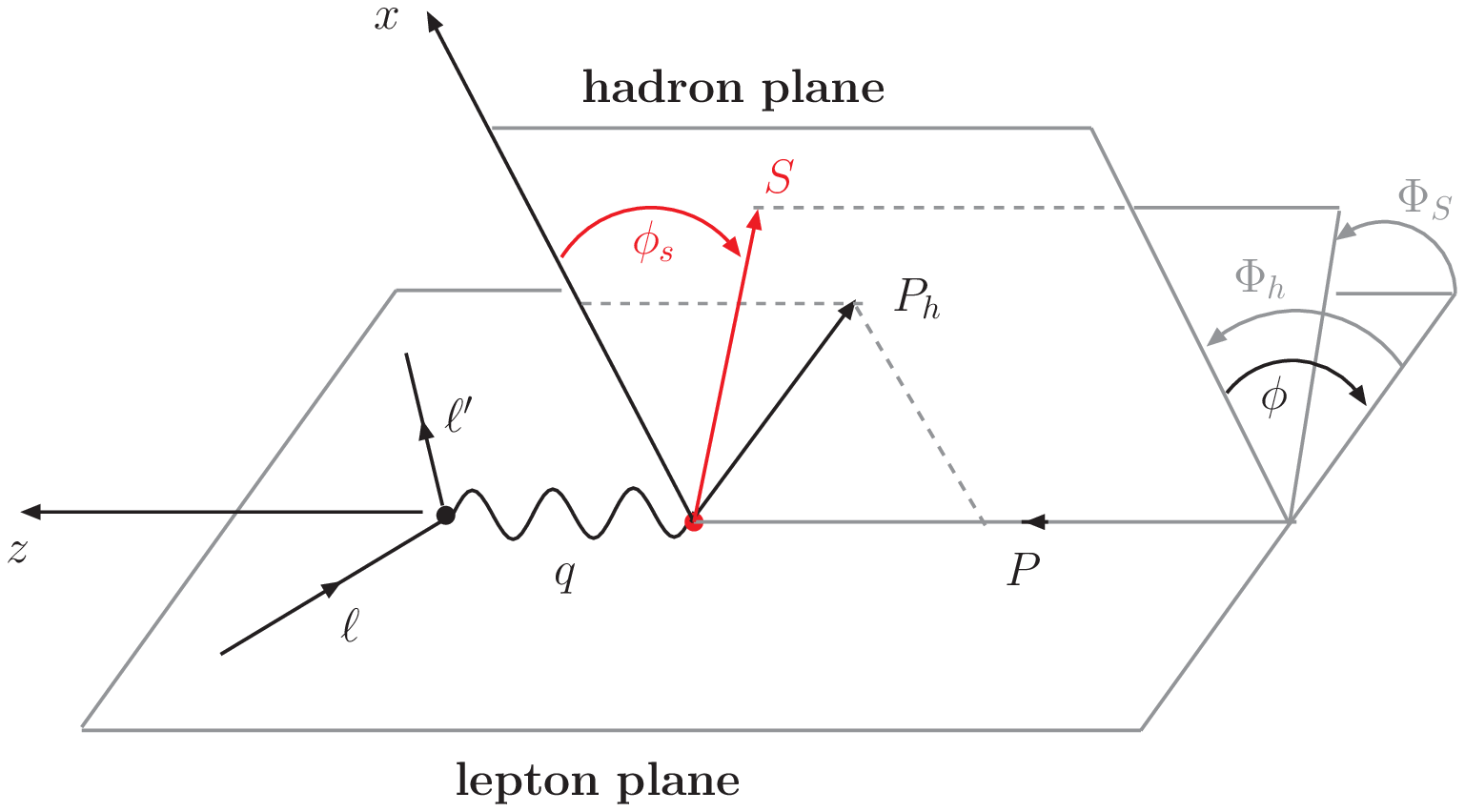,width=3.7in}
\caption{Kinematics of the SIDIS process in hadron frame.}
\label{frame}
\eef

The single transverse-spin asymmetry is defined as
\ben
A_N=\frac{\sigma(s_\perp)-\sigma(-s_\perp)}{\sigma(s_\perp)+\sigma(-s_\perp)}
=\frac{d\Delta\sigma(s_\perp)}{dx_B dy dz_h dP_{h\perp}^2 d\phi}\left/
\frac{d\sigma}{dx_B dy dz_h dP_{h\perp}^2 d\phi}\right..
\label{AN}
\een
In the following subsections, we will first review the unpolarized cross section at leading order, and then derive the single-transverse polarized cross sections, $\Delta\sigma(s_\perp)$.
\subsection{Unpolarized cross section}
The unpolarized differential SIDIS cross section may be calculated from the formula
\ben
\frac{d\sigma}{dx_B dy dz_h dP_{h\perp}^2 d\phi}=\frac{\pi \alpha_{em}^2 y}{Q^4}
L_{\mu\nu}(\ell,\,q)W^{\mu\nu}(P,\,q,\, P_h),
\label{LW}
\een
where $L_{\mu\nu}$ and $W^{\mu\nu}$ are the leptonic and hadronic tensors, respectively. The leptonic tensor is given by
\ben
L_{\mu\nu}(\ell,\,q)=2\left(\ell_\mu \ell'_\nu+\ell'_\nu \ell_\mu-g_{\mu\nu}Q^2/2\right).
\een
The hadronic tensor has the following expression in QCD:
\ben
W^{\mu\nu}(P,\,q,\, P_h)=\frac{1}{4z_h}\sum_{X}\int\frac{d^4\xi}{(2\pi)^4}e^{iq\cdot \xi}\langle P|J_\mu(\xi)|X\,P_h\rangle\langle X\,P_h|J_\nu(0)|P\rangle,
\een
where $J^\mu$ is the quark electromagnetic current and $X$ represents all other final-state hadrons other than the observed open charm meson $h$.

The hadronic tensor can be further decomposed in terms of five parity and current conserving tensors ${\cal V}_i^{\mu\nu}$ \cite{sidis}:
\ben
W^{\mu\nu}=\sum_{i=1}^5 {\cal V}_i^{\mu\nu} W_i,
\een
where the $W_i$ are structure functions which may be projected out from $W^{\mu\nu}$ by $W_i=W_{\rho\sigma}\tilde{{\cal V}}_i^{\rho\sigma}$, with the corresponding inverse tensors $\tilde{{\cal V}}_i$. Both ${\cal V}_i$ and $\tilde{{\cal V}}_i$ can be constructed from four orthonormal basis vectors:
\ben
T^\mu&=&\frac{1}{Q}\left( q^\mu+2x_B P^\mu \right),\nnu
X^\mu&=&\frac{1}{q_\perp}\left[ \frac{P_h^\mu}{z_h}-q^\mu-\left( 1+\frac{q_\perp^2+m_h^2/z_h^2}{Q^2} \right)x_B P^\mu \right],\nnu
Y^\mu&=&\epsilon^{\mu\nu\rho\sigma}Z_\nu X_\rho T_\sigma,\nnu
Z^\mu&=&-\frac{q^\mu}{Q},
\een
with normalization $T^2=1$ and $X^2=Y^2=Z^2=-1$, which are reduced to those in  \cite{sidis} when $m_h=0$. The tensor ${\cal V}_5$ does not contribute to the cross section when it is contracted with a symmetric $L_{\mu\nu}$, the other four tensors and their inverse are given as \cite{sidis}:
\ben
&&{\cal V}^{\mu\nu}_1=X^\mu X^\nu+Y^\mu Y^\nu, \qquad
{\cal V}^{\mu\nu}_2=g^{\mu\nu}+Z^\mu Z^\nu, \qquad \nnu
&&{\cal V}^{\mu\nu}_3=T^\mu X^\nu+T^\nu X^\mu, \qquad
{\cal V}^{\mu\nu}_4=X^\mu X^\nu-Y^\mu Y^\nu, \\
&&\tilde{{\cal V}}^{\mu\nu}_1=\frac{1}{2}\left(2T^\mu T^\nu+X^\mu X^\nu+Y^\mu Y^\nu\right), \qquad
\tilde{{\cal V}}^{\mu\nu}_2=T^\mu T^\nu, \qquad \nnu
&&\tilde{{\cal V}}^{\mu\nu}_3=-\frac{1}{2}\left(T^\mu X^\nu+T^\nu X^\mu\right), \qquad
\tilde{{\cal V}}^{\mu\nu}_4=\frac{1}{2}\left(X^\mu X^\nu-Y^\mu Y^\nu\right).
\een
The contraction of $L_{\mu\nu}$ and ${\cal V}^{\mu\nu}_i$ leads to various angular distributions.  Let ${\cal A}_i=L_{\mu\nu}{\cal V}_i^{\mu\nu}/Q^2$, we have  
\ben
{\cal A}_1=1+\cosh^2\psi, \qquad
{\cal A}_2=-2, \qquad
{\cal A}_3=-\cos\phi \sinh{2\psi}, \qquad
{\cal A}_4=\cos{2\phi} \sinh^2{\psi}.
\een
We can then write the cross section in Eq.~(\ref{LW}) as
\ben
\frac{d\sigma}{dx_B dy dz_h dP_{h\perp}^2 d\phi}=\frac{\pi\alpha_{em}^2 y}{Q^2}
\sum_{i=1}^4 {\cal A}_iW_i.
\een

At large $P_{h\perp}\sim Q$, the collinear factorization is expected to be valid, and $W_i$ can be factorized into a convolution of the parton distribution function, the fragmentation function for the produced $D$ meson, and a short-distance partonic hard part. The lowest-order (LO) contribution to the partonic hard part comes from the photon-gluon fusion subprocess $\gamma^*+g\to Q(p_c)+\bar{Q}(p_{\bar{c}})$, which gives the leading order cross section as
\ben
\frac{d\sigma}{dx_B dy dz_h dP_{h\perp}^2 d\phi}
&=&
\sigma_0 \int_{x_{min}}^1\frac{dx}{x} \int \frac{dz}{z}\,
G(x)D(z)\,
\delta\left(\frac{P_{h\perp}^2}{z_h^2}
           -\frac{(1-\hat{x})(1-\hat{z})}{\hat{x}\hat{z}}Q^2
           +\hat{z}^2 m_c^2\right)
\left(\frac{1}{2}\right)\sum_{i=1}^4{\cal A}_i\hat{W}_i,
\label{unpolarized}
\een
where $\sigma_0=e_c^2\alpha_{em}^2\alpha_s y/(8\pi z_h^2 Q^2)$, $\hat{x}=x_B/x$, $\hat{z}=z_h/z$, and $e_c$ and $m_c$ are the fractional charge and mass of the charm quark, respectively. The $P_{h\perp}^2/z_h^2$ in the $\delta$-function could be replaced by $q_\perp^2$, and the $1/2$ is the color factor.  In Eq.~(\ref{unpolarized}) $G(x)$ is the unpolarized gluon distribution function with gluon momentum fraction $x$, and $D(z)$ is the fragmentation function for the charm quark to become a $D$ meson with $z=P\cdot P_h/P\cdot p_c$. We have suppressed the dependence on the factorization and renormalization scales for simplicity. We used $P_{h\perp}\approx z p_{c\perp}$ inside the $\delta$-function, which fixes the $z$ integration. The lower limit of $x$ integration $x_{min}$ is given by:
\ben
x_{min} = \left\{ 
\begin{array}{ll}
x_B\left[1+\frac{P_{h\perp}^2+m_c^2}{z_h(1-z_h)Q^2}\right],
& \quad \mbox{if } z_h+\sqrt{z_h^2+\frac{P_{h\perp}^2}{m_c^2}}\geq 1;\\ 
\\
x_B\left[1+\frac{2m_c^2}{Q^2}\left(1+\sqrt{1+\frac{P_{h\perp}^2}	 
{z_h^2m_c^2}}\right)\right],
& \quad \mbox{if } z_h+\sqrt{z_h^2+\frac{P_{h\perp}^2}{m_c^2}}\leq 1.\\
\end{array} \right.
\label{xmin}
\een
The short-distance parts $\hat{W}_i$ are calculated from the photon-gluon scattering and are given by
\ben
\hat{W}_1
&=&
2\left[\frac{\hat{u}}{\hat{t}}+\frac{\hat{t}}{\hat{u}}
      -\frac{2\hat{s}Q^2}{\hat{t}\hat{u}}
      +\frac{4\hat{x}^2\hat{s}}{Q^2}\right]
+4m_c^2\left[\frac{Q^2-2\hat{t}}{\hat{t}^2}+\frac{Q^2-2\hat{u}}{\hat{u}^2}
      -\frac{2\hat{x}^2}{Q^2}\left(\frac{\hat{u}}{\hat{t}}
      +\frac{\hat{t}}{\hat{u}}+2\right)\right]
-8m_c^4\left[\frac{1}{\hat{t}}+\frac{1}{\hat{u}}\right]^2,\nnu
\hat{W}_2
&=&
\frac{16\hat{x}^2}{Q^2}\left[\hat{s}-m_c^2\left(\frac{\hat{u}}{\hat{t}}
      +\frac{\hat{t}}{\hat{u}}+2\right)\right],\nnu
\hat{W}_3
&=&
4\hat{x}\hat{z}\frac{q_\perp}{Q}(\hat{u}-\hat{t})
\left[\frac{\hat{s}-Q^2}{\hat{t}\hat{u}}
     -2m_c^2\left(\frac{1}{\hat{t}}+\frac{1}{\hat{u}}\right)^2 \right],
\nnu
\hat{W}_4
&=&
8\hat{z}^2q_\perp^2\left[\frac{Q^2}{\hat{t}\hat{u}}
+m_c^2\left(\frac{1}{\hat{t}}+\frac{1}{\hat{u}}\right)^2 \right],
\label{WLO}
\een
where $\hat{s}, \hat{t}, \hat{u}$ are defined at the partonic level as
\ben
\hat{s}\equiv (xP+q)^2=\frac{1-\hat{x}}{\hat{x}}Q^2,
\qquad 
\hat{t}\equiv (p_c-q)^2-m_c^2=-\frac{1-\hat{z}}{\hat{x}}Q^2, 
\qquad 
\hat{u}\equiv (xP-p_c)^2-m_c^2=-\frac{\hat{z}}{\hat{x}}Q^2\, ,
\label{stu}
\een
which are different from some definitions used in the literature.  We found that 
this definition makes the expression of $\hat{W}_i$ for massive quark production simpler.  Taking $m_c=0$ in Eqs.~(\ref{WLO}) and (\ref{stu}), one recovers the results for the production of massless quark derived in \cite{Mendez:1978zx, Koike}.

\subsection{Twist-three polarized cross section}
We now proceed to derive the single transverse-spin dependent cross section by applying the same method developed in Refs.~\cite{qiu,Kouvaris:2006zy}. When both physically observed scales $Q, P_{h\perp}\gg \Lambda_{\rm QCD}$, the spin-dependent cross section for $D$-meson production is expected to be factorized in terms of twist-three transverse-spin dependent tri-gluon correlation function \cite{Qiu:1990cu},
\ben
d\Delta\sigma(s_\perp)\propto \frac{1}{2 S_{ep}}\int dz D(z) 
\int dx_1 dx_2 {\cal T}_G(x_1,x_2)\ i \epsilon^{\rho s_\perp n\bar{n}}
\lim_{k_\perp\to 0}\frac{\partial}{\partial k_\perp^\rho}H(x_1,x_2,k_\perp),
\label{Dsig_form}
\een
where $1/2S_{ep}$ is the flux factor and $\epsilon^{\rho s_\perp n\bar{n}}=\epsilon^{\rho\sigma\mu\nu}s_{\perp\sigma}n_\mu\bar{n}_\nu$,
\ben
{\cal T}_G(x_1,x_2)=\int \frac{P^+dy_1^- dy_2^-}{2\pi}e^{ix_1P^+y_1^- + i (x_2-x_1)P^+y_2^-}d_{\alpha\beta}\langle P,s_\perp| A^{\alpha}(0)\left[ \epsilon^{s_\perp\sigma n\bar{n}}F_\sigma^{~ +}(y_2^-)\right] A^{\beta}(y_1^-)|P, s_\perp\rangle,
\een
where $d_{\alpha\beta}=-g_{\alpha\beta}+\bar{n}_\alpha n_\beta+\bar{n}_\beta n_\alpha$. ${\cal T}_G(x_1,x_2)$ is related to the tri-gluon correlation function through $T_G(x,x)=x{\cal T}_G(x,x)$. Since ${\cal T}_G(x_1,x_2)$ is real, we need an imaginary part of the hard-scattering function $H(x_1, x_2, k_\perp)$ to contract with $i \epsilon^{\rho s_\perp n\bar{n}}$ in order to obtain a real $\Delta\sigma(s_\perp)$. This imaginary part comes from the interference between a real part of scattering amplitude with a single gluon initial state and an imaginary part of the partonic scattering amplitude with an extra gluon, see Fig. \ref{phase}. Technically, the imaginary part, or the phase, ``$i$'', arises when the virtual momentum integral of the extra gluon is evaluated by the residue of an unpinched pole from a propagator in the amplitude with an extra gluon.  Such propagator is indicated by the one marked with a short bar in the diagrams in Fig.~\ref{twist3_LO}. 
\bef
\psfig{file=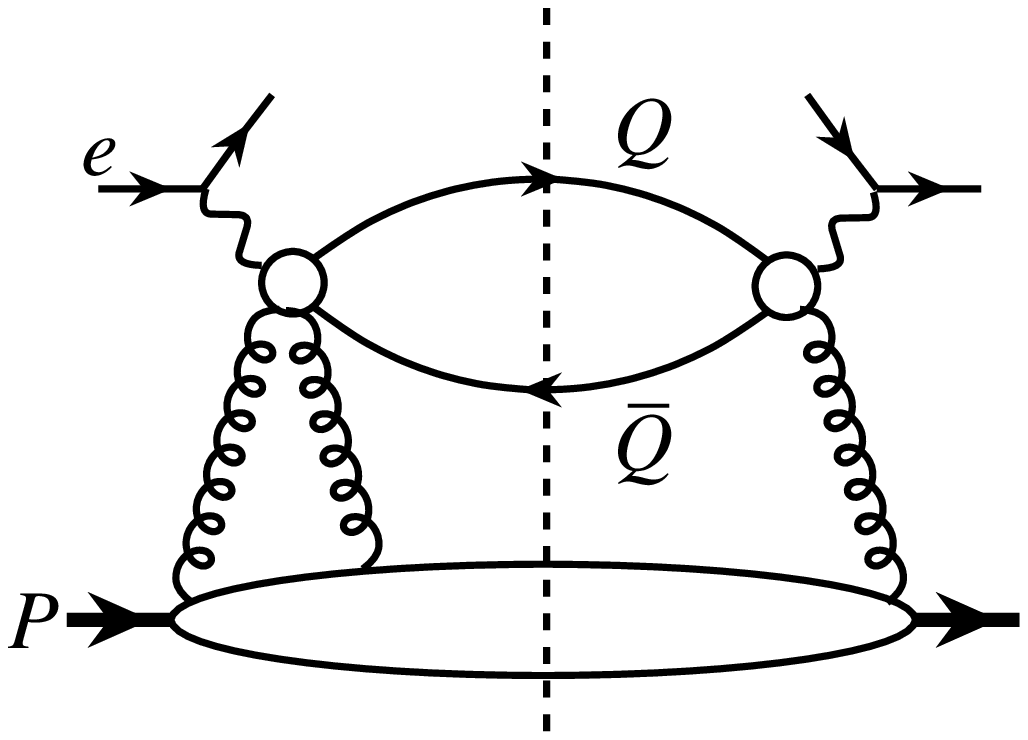,width=1.4in}
\caption{A typical diagram that gives a non-vanishing contribution to the SSA.}
\label{phase}
\eef

There are a total of eight partonic diagrams contributing to the twist-three polarized cross sections, $\Delta\sigma(s_\perp)$. Four of them are shown in Fig.~\ref{twist3_LO}, and the other four are obtained by attaching the extra gluon in the same way on the right side of the final-state cut. When the extra gluon is attached to the left side of the final-state cut, as shown in Fig.~\ref{twist3_LO}, the phase from the propagator marked by the bar arises effectively as
\ben
\frac{1}{\left(p_c-(x_2-x_1)P-k_\perp\right)^2-m_c^2+i\epsilon}&=&\frac{1}{2P\cdot p_c}\frac{1}{x_1-x_2+v_1\cdot k_\perp+i\epsilon}+{\cal O}(k_\perp^2)\nnu
&\rightarrow& \frac{-i\pi}{2P\cdot p_c}\delta(x_1-x_2+v_1\cdot k_\perp),
\een 
to fix the virtual loop momentum fraction $x_1=x_2-v_1\cdot k_\perp$ with $v_1^\mu=-2p_c^\mu/2P\cdot p_c$. 
\bef
\psfig{file=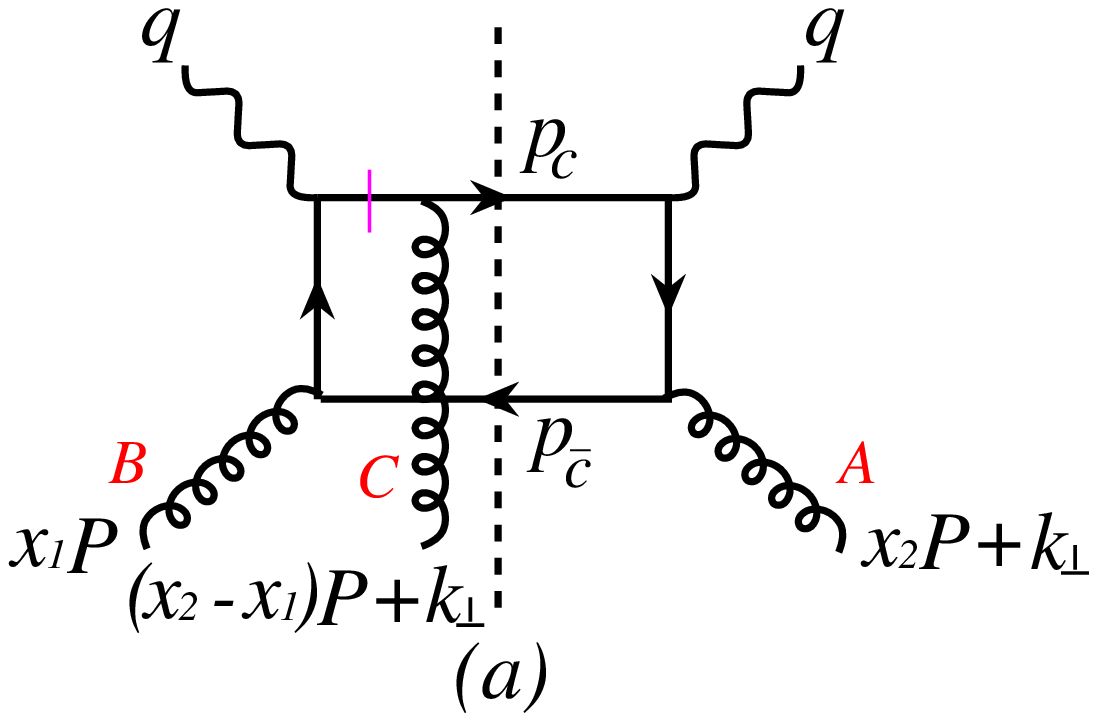,width=1.65in}
\hskip 0.1in
\psfig{file=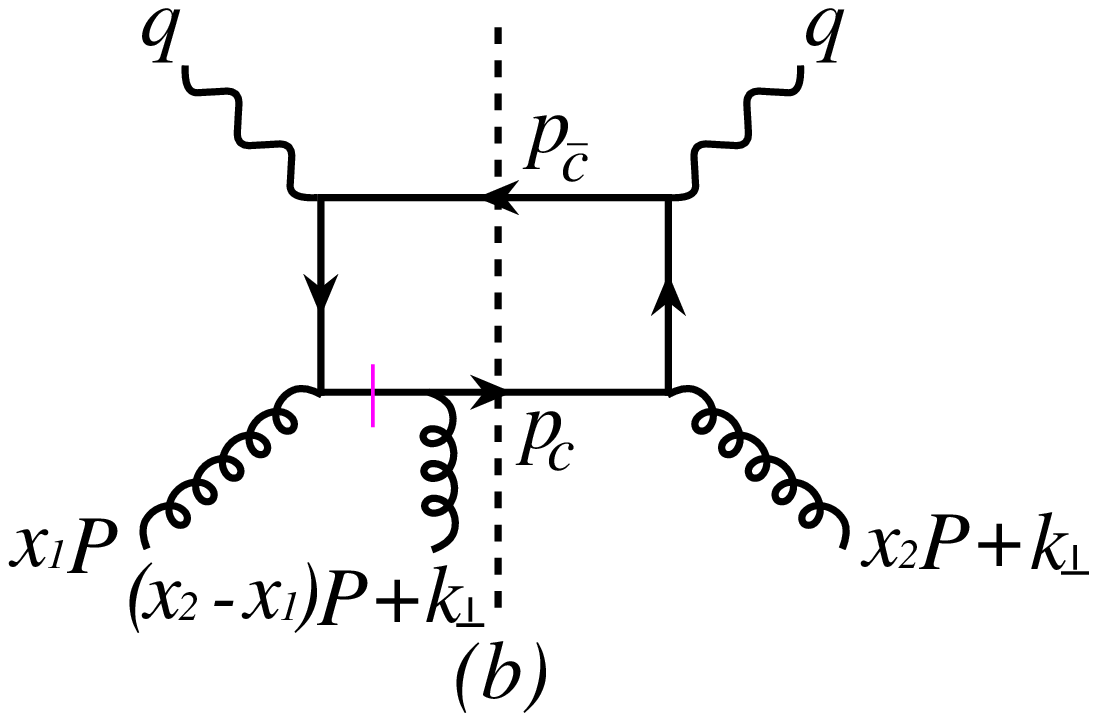,width=1.65in}
\hskip 0.1in
\psfig{file=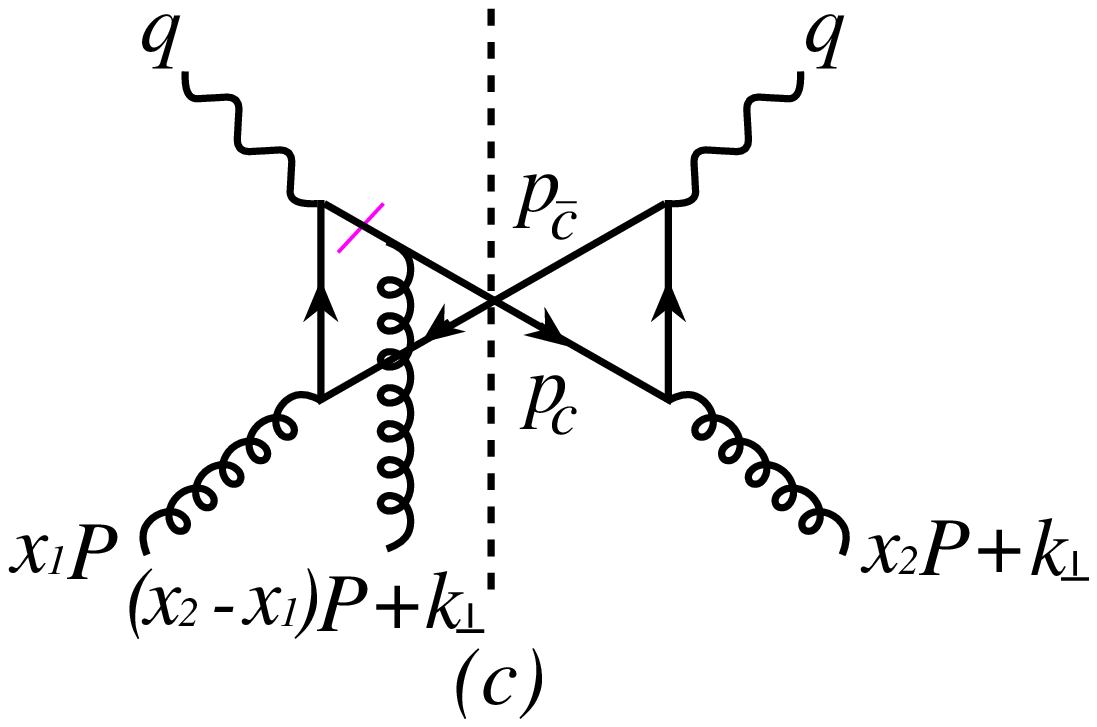,width=1.65in}
\hskip 0.1in
\psfig{file=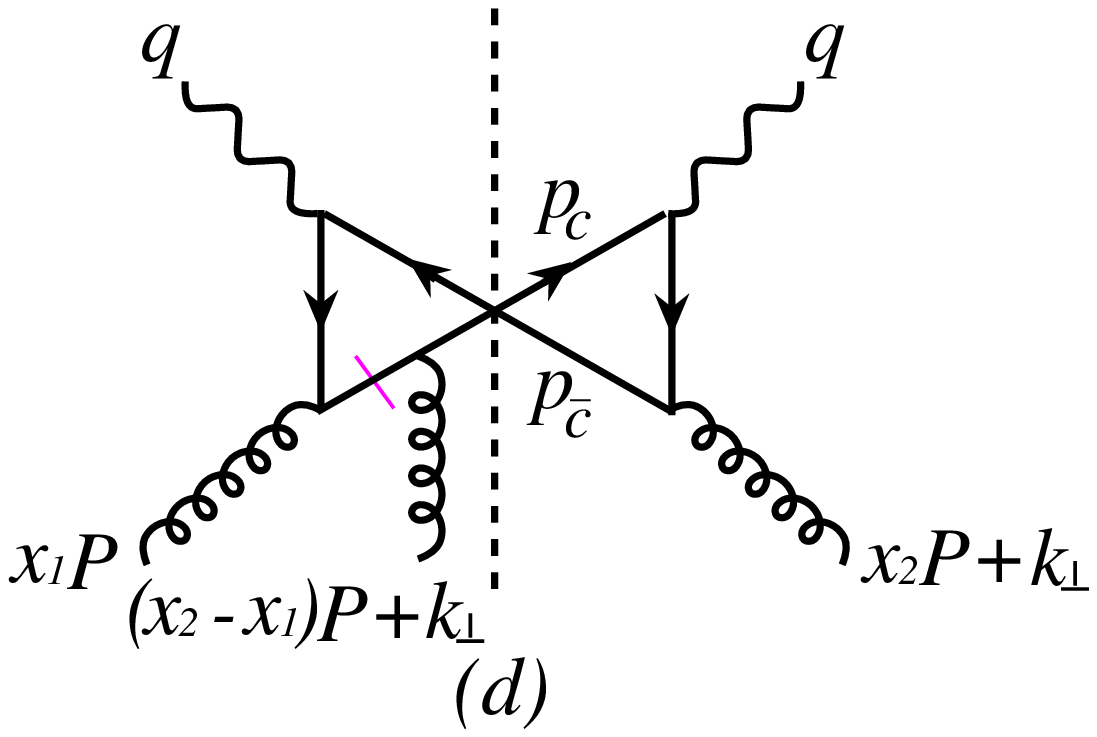,width=1.65in}
\caption{Feynman diagrams that give the twist-three contribution to the spin-dependent cross section. The short bar indicates the propagator that produces the pole.  The letters, $A,B$ and $C$, represent the color of the initial-state gluons.}
\label{twist3_LO}
\eef
On the other hand, the on-shell condition associated with the unobserved anti-charm quark fixes the momentum fraction of the active initial-state gluon as
\ben
\delta(p_{\bar{c}}^2-m_c^2)&=&\delta\left((x_2 P+k_\perp+q-p_c)^2-m_c^2\right)\nnu
&=&\frac{1}{2P\cdot (q-p_c)}\delta(x_2-x-v_2\cdot k_\perp),
\een
where terms at ${\cal O}(k_\perp^2)$ and higher are neglected and 
\ben
x=-\frac{(q-p_c)^2-m_c^2}{2P\cdot (q-p_c)}, \quad\quad 
v_2^\mu=\frac{2p_c^\mu}{2P\cdot (q-p_c)}.
\een
When the extra gluon is attached to the right hand side of the cut, the phase arises as
\ben
\frac{1}{\left(p_c+(x_2-x_1)P+k_\perp\right)^2-m_c^2-i\epsilon}&=&\frac{1}{2P\cdot p_c}\frac{1}{x_2-x_1-v_1\cdot k_\perp-i\epsilon}+{\cal O}(k_\perp^2)\nnu
&\rightarrow& \frac{i\pi}{2P\cdot p_c}\delta(x_2-x_1-v_1\cdot k_\perp),
\een
and  the on-shell condition of the unobserved anti-charm quark gives
\ben
\delta(p_{\bar{c}}^2-m_c^2)=\frac{1}{2P\cdot (q-p_c)}\delta(x_1-x),
\een
which has no $k_\perp$-dependence.

Applying the so-called ``master formula'' in Ref.~\cite{Kouvaris:2006zy}, we have from Eq.~(\ref{Dsig_form}) the following general expression:
\ben
&&\lim_{k\perp\to 0}\frac{\partial}{\partial k_\perp^\rho}\int dx_1\int dx_2\, {\cal T}_G(x_1,x_2) 
\left[ H_L(x_1, x_2, k_\perp)\delta(x_1-x_2+v_1\cdot k_\perp)\delta(x_2-x-v_2\cdot k_\perp) \right. \nnu
&&\left. -H_R(x_1, x_2, k_\perp)\delta(x_2-x_1-v_1\cdot k_\perp)\delta(x_1-x)\right]\nnu
&&=(v_2-v_1)^\rho H_L(x,x,0)\frac{d}{dx}\left(\frac{T_G(x,x)}{x}\right)+\frac{T_G(x,x)}{x}
\nnu
&&\times\lim_{k_\perp\to 0}\frac{\partial}{\partial k_\perp^\rho}
\left[ H_L(x+(v_2-v_1)\cdot k_\perp, x+v_2\cdot k_\perp, k_\perp)-H_R(x, x+v_1\cdot k_\perp, k_\perp)\right],
\label{master}
\een
where we have already used the facts that $H_L(x,x,0)=H_R(x,x,0)$ and $T_G(x,x)=x{\cal T}_G(x,x)$.  The fact that Eq.~(\ref{master}) depends only on the diagonal part of the tri-gluon correlation function, $T_G(x_1,x_2)$, with $x_1=x_2=x$ is a consequence of that the photon-gluon fusion subprocess at this order has only the so-called ``soft-pole'' contribution to the SSA \cite{qiu,UnifySSA}.  Therefore, the measurement of the SSA in $D$-meson production in SIDIS is a direct measurement of the tri-gluon correlation function, $T_G(x,x)$.  

In terms of $\hat{s},\hat{t},\hat{u}$ defined in the previous subsection, we have
\ben
v_1^\mu=\frac{2x}{\hat{u}}p_c^\mu, 
\qquad 
v_2^\mu=-\frac{2x}{\hat{t}} p_c^\mu,
\qquad 
\left(v_2-v_1\right)^\mu=-\frac{2x}{\hat{t}}\left(1+\frac{\hat{t}}{\hat{u}}\right)p_c^\mu.
\een
Using Eqs. (\ref{Dsig_form}), (\ref{master}), and adding the contributions from the eight diagrams together, we find the final expression for the fully differential single-transverse-spin-dependent cross section:
\ben
\frac{d\Delta\sigma(s_\perp)}{dx_B dy dz_h dP_{h\perp}^2 d\phi}
&=&
\sigma_0\int_{x_{min}}^1 dx\int \frac{dz}{z}D(z)
\delta\left(\frac{P_{h\perp}^2}{z_h^2}
           -\frac{(1-\hat{x})(1-\hat{z})}{\hat{x}\hat{z}}Q^2
           +\hat{z}^2 m_c^2\right)
\left(\frac{1}{4}\right) \nnu
&\times&
\left[\epsilon^{P_h s_\perp n \bar{n}}
\left(\frac{\sqrt{4\pi\alpha_s}}{z\hat{t}}\right)
\left(1+\frac{\hat{t}}{\hat{u}}\right)\right]
\sum_{i=1}^{4} {\cal A}_i 
\left[-x\frac{d}{d x}\left(\frac{T_G(x,x)}{x}\right)\hat{W}_i+\left(\frac{T_G(x,x)}{x}\right)\hat{N}_i\right],
\label{polarized}
\een
where $1/4$ is the color factor, $T_G(x, x)$ is the tri-gluon correlation function defined in Eq.~(\ref{TG_correlation}), $\hat{W}_i$ are given in Eq.~(\ref{WLO}), and the hard parts for the ``non-derivative'' term, 
$\hat{N}_i$, are given by
\ben
\hat{N}_1
&=&
4\left[\frac{2m_c^2-Q^2}{\hat{t}\hat{u}}+\frac{6\hat{x}^2}{Q^2}\right]\left[\left(\hat{s}-Q^2\right)
     -2m_c^2\left(\frac{\hat{u}}{\hat{t}}
                 +\frac{\hat{t}}{\hat{u}}+2\right)\right],
\nnu
\hat{N}_2
&=&
\frac{16\hat{x}^2}{Q^2}\left[\left(\hat{s}-Q^2\right)
     -2m_c^2\left(\frac{\hat{u}}{\hat{t}}
                 +\frac{\hat{t}}{\hat{u}}+2\right)\right],
\nnu
\hat{N}_3
&=&
\frac{2Q}{\hat{z}q_\perp}\left(\hat{u}-\hat{t}\right)
\left[\left(\frac{4\hat{z}^2q_\perp^2}{\hat{t}\hat{u}}
           -\frac{1}{Q^2+\hat{s}}\right)
\left(2m_c^2\left(\frac{1}{\hat{t}}+\frac{1}{\hat{u}}\right)
-\frac{Q^2-\hat{s}}{Q^2+\hat{s}}\right)-2\hat{z}q_\perp^2\right],
\nnu
\hat{N}_4
&=&
8\left[2\hat{z}q_\perp^2-\frac{\hat{t}\hat{u}}{Q^2+\hat{s}}\right]
\left[\frac{Q^2}{\hat{t}\hat{u}}+m_c^2\left(\frac{1}{\hat{t}}
     +\frac{1}{\hat{u}}\right)^2\right].
\label{NLO}
\een
Eq.~(\ref{polarized}) is our main result for the leading order twist-three $T_G(x,x)$ contribution to the fully differential polarized cross section, $\Delta\sigma(s_\perp)$, of $D$-meson production in SIDIS.  The single transverse-spin asymmetry for the $D$-meson production in SIDIS is obtained by substituting Eqs.~(\ref{unpolarized}) and (\ref{polarized}) into Eq.~(\ref{AN}).

Similar to the twist-three contributions to the SSAs generated by the fermionic quark-gluon correlation function, $T_F(x,x)$, the gluonic twist-three contribution to the SSA of $D$-meson production in Eq.~(\ref{polarized}) has both the ``derivative'' and ``non-derivative'' terms, a unique feature of twist-three contribution.  It was found that the fermionic ``non-derivative'' and ``derivative'' terms can be combined into a simple form $T_F(x,x)-xT'_F(x,x)$ \cite{Kouvaris:2006zy,Koike:proof}.  However, from Eqs.~(\ref{WLO}) and (\ref{NLO}), it is clear that $\hat{W}_i\neq \hat{N}_i$. That is, our explicit calculation shows that such a simple combination does not hold for contributions from tri-gluon correlation function $T_G(x,x)$, and is not universal.

We close this section by a discussion on the calculation of the color factor $1/4$ in Eq.~(\ref{polarized}). The color factor in Eq.~(\ref{polarized}) depends on how colors of the three gluon fields in the matrix element of the tri-gluon correlation function in Eq.~(\ref{TG_correlation}) are neutralized.  If the color of the operator, $F^A(0)\,F^C(y_2^-)\,F^B(y_1^-)$, in Eq.~(\ref{TG_correlation}) with the Lorentz indices suppressed, is neutralized by $({\cal F}^C)_{AB}=-if^{CAB}$ with $f^{CAB}$ the fully antisymmetric structure constant of the color SU(3) group, we refer to this tri-gluon correlation function as $T_G(x,x)$ as expressed in Eq.~(\ref{TG_correlation}).  The corresponding color factor for the partonic part in Eq.~(\ref{polarized}) is calculated by contracting the color indices of the Feynman diagrams in Fig.~\ref{twist3_LO} with $\frac{i}{N(N^2-1)}\,f^{ABC}$, which gives the color factor $1/4$ in Eq.~(\ref{polarized}).  On the other hand, if the color of the operator, $F^A(0)\,F^C(y_2^-)\,F^B(y_1^-)$, is neutralized by $({\cal D}^C)_{AB}=d^{CAB}$, which is symmetric under the interchange of any two indices, we have a different tri-gluon correlation function and we refer it as $\widetilde{T}_G(x,x)$, which has the same expression as $T_G(x,x)$ except the difference in the contraction of the gluon color.  The color factor for the corresponding partonic hard part is calculated by contracting the color indices of the same Feynman diagrams with $\frac{N}{(N^2-1)(N^2-4)}\,d^{ABC}$, which also gives the color factor $1/4$.  

That is, there could be two tri-gluon correlation functions depending on how the colors of the three gluon fields are neutralized \cite{Yuan:private}.  We find that both correlation functions are gauge invariant after inserting necessary gauge link in the adjoint representation between the gluon field strengths in the matrix element in Eq.~(\ref{TG_correlation}), and they contribute to the SSAs with the same partonic hard parts, and potentially, different color factors \cite{Kang:hadron}.  Since the color factors are the same in our case, adding the potential contribution from $\widetilde{T}_G(x,x)$ is to replace the $T_G(x,x)$ in Eq.~(\ref{polarized}) by $T_G(x,x)+\widetilde{T}_G(x,x)$.

We noticed that all Feynman diagrams for producing a charm quark at this order in Fig.~\ref{twist3_LO} have the same color structure, ${\rm Tr}[T^A T^C T^B]=(d^{ACB}+if^{ACB})/4$, which leads to the overall color factor 1/4 for the contributions from both $T_G(x,x)$ and $\widetilde{T}_G(x,x)$.  However, for the SSAs of producing a $\bar{D}$ meson, which is fragmented from an anticharm quark, both the partonic part and the antisymmetric part of the color structure change sign, while the symmetric part of the color structure is unchanged.  Therefore, the SSAs for $\bar{D}$-meson production in the SIDIS at the leading order have the same functional form as that for the $D$-meson produciton except the sum of the tri-gluon correlation functions, $T_G(x,x)+\widetilde{T}_G(x,x)$, is replaced by $T_G(x,x)-\widetilde{T}_G(x,x)$.  That is, by comparing the 
SSAs for producing $D$ and $\bar{D}$ mesons in SIDIS, we could gain valuable information on both tri-gluon correlation functions.  However, the relation could be complicated in the $D$-meson production in hadronic collisions due to the additional color flow from the other colliding hadron \cite{Kang:hadron}.

We also notice that the correlation function, $T_G(x,x)$, with the color neutralized by $({\cal F}^C)_{AB}=-if^{CAB}$ could potentially be related to the spin-dependent TMD gluon distribution, or the gluonic Sivers function\cite{Anselmino:2004nk}, since the middle gluon field strength of the operator, $F^A(0)\,F^C(y_2^-)\,F^B(y_1^-)$, could be related to the gauge link in the adjoint representation that is needed to define the TMD gluon distribution.  Without knowing the size and sign of either tri-gluon correlation functions, we will treat them as one combined tri-gluon correlation function, labeled by $T_G(x,x)$, in the following numerical estimates of the SSAs.

\section{Phenomenology}
\label{numerical}

In this section we first evaluate the inclusive $D$-meson production rate at large $P_{h\perp}$ in SIDIS. We then propose a simple model for tri-gluon correlation function $T_G(x, x)$ and estimate the size of SSA for the $D$-meson production in SIDIS.

The charm meson's transverse momentum, $P_{h\perp}$, is chosen to be along the $x$-direction in the {\it hadron frame}, and therefore,
\ben
\epsilon^{P_h s_\perp n \bar{n}}=-P_{h\perp}\sin\phi_s.
\een
The fully differential cross sections in Eqs.~(\ref{unpolarized}) and (\ref{polarized}) can be decomposed in terms of the independent angular distributions as follows,
\ben
\frac{d\sigma}{dx_B dy dz_h dP_{h\perp}^2 d\phi}&=&\sigma_0^U+\sigma_1^U\cos{\phi}+\sigma_2^U \cos{2\phi}, \nnu
\frac{d\Delta\sigma}{dx_B dy dz_h dP_{h\perp}^2 d\phi}&=&\sin{\phi_s}\left(\Delta\sigma_0+\Delta\sigma_1\cos{\phi}+\Delta\sigma_2 \cos{2\phi}\right).
\label{angulardis}
\een
Before evaluating the SSA, we first estimate the $D$-meson production rate in the unpolarized SIDIS by using our LO formula in Eq.~(\ref{unpolarized}). 

For the following numerical evaluations we use CTEQ6L parton distribution functions \cite{Pumplin:2002vw}, and charm-to-$D$ fragmentation functions from Ref.~\cite{Kneesch:2007ey}.  We choose the factorization scale to be equal to the renormalization scale and set $\mu=\sqrt{Q^2+m_c^2+P_{h\perp}^2}$ with charm quark mass $m_c=1.3$~GeV. In the following plots, we choose two sets of kinematic variables. The first one is $S_{ep}=300$ GeV$^2$, $x_B=0.01$ and $Q=1$ GeV, which is close to the COMPASS kinematics.  The other is $S_{ep}=2500$ GeV$^2$, $x_B=0.01$ and $Q=4$ GeV, which is more relevant to the planned eRHIC experiment  \cite{Deshpande:2005wd}. 
\bef
\psfig{file=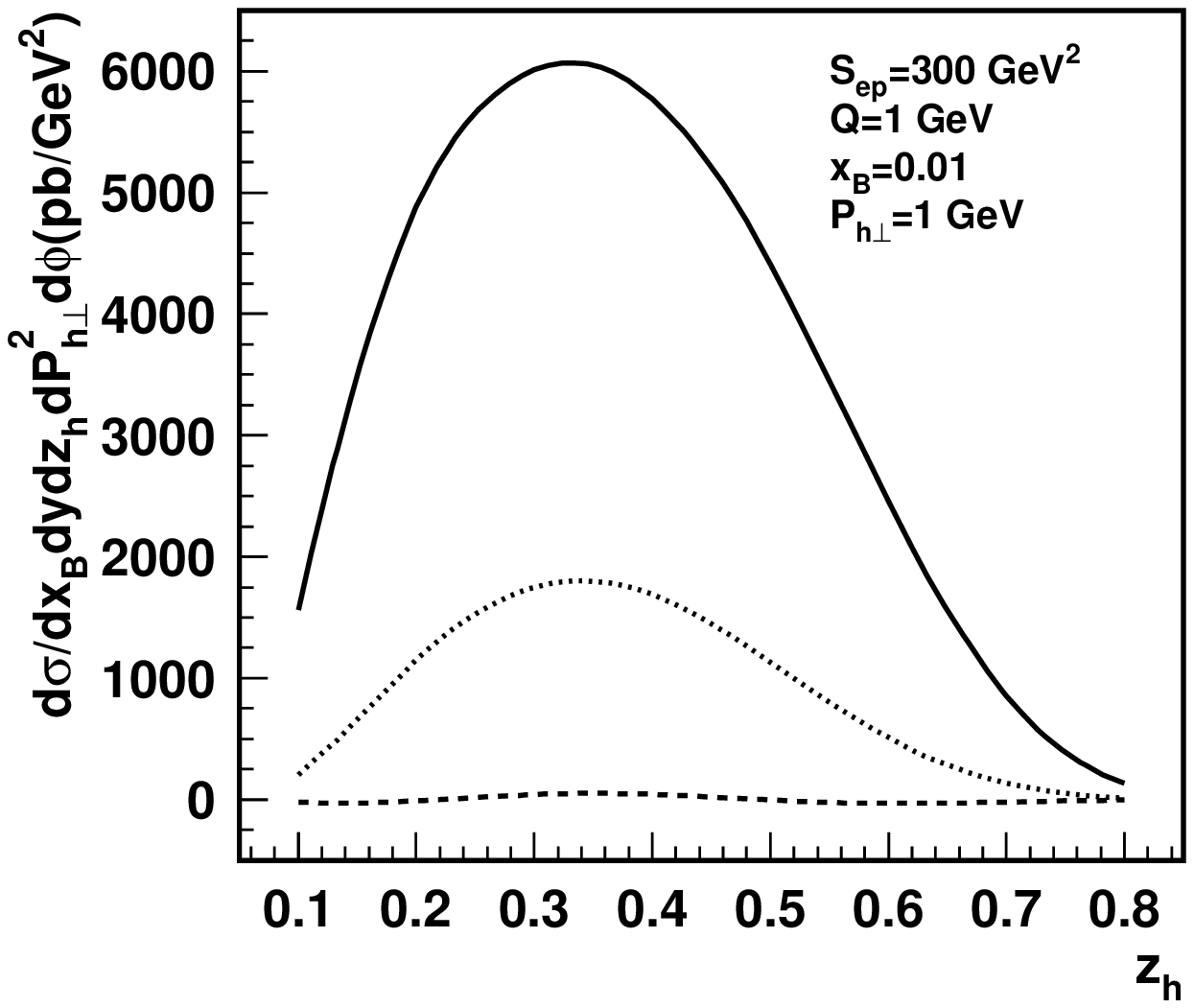,height=2.5in}
\hskip 0.2in
\psfig{file=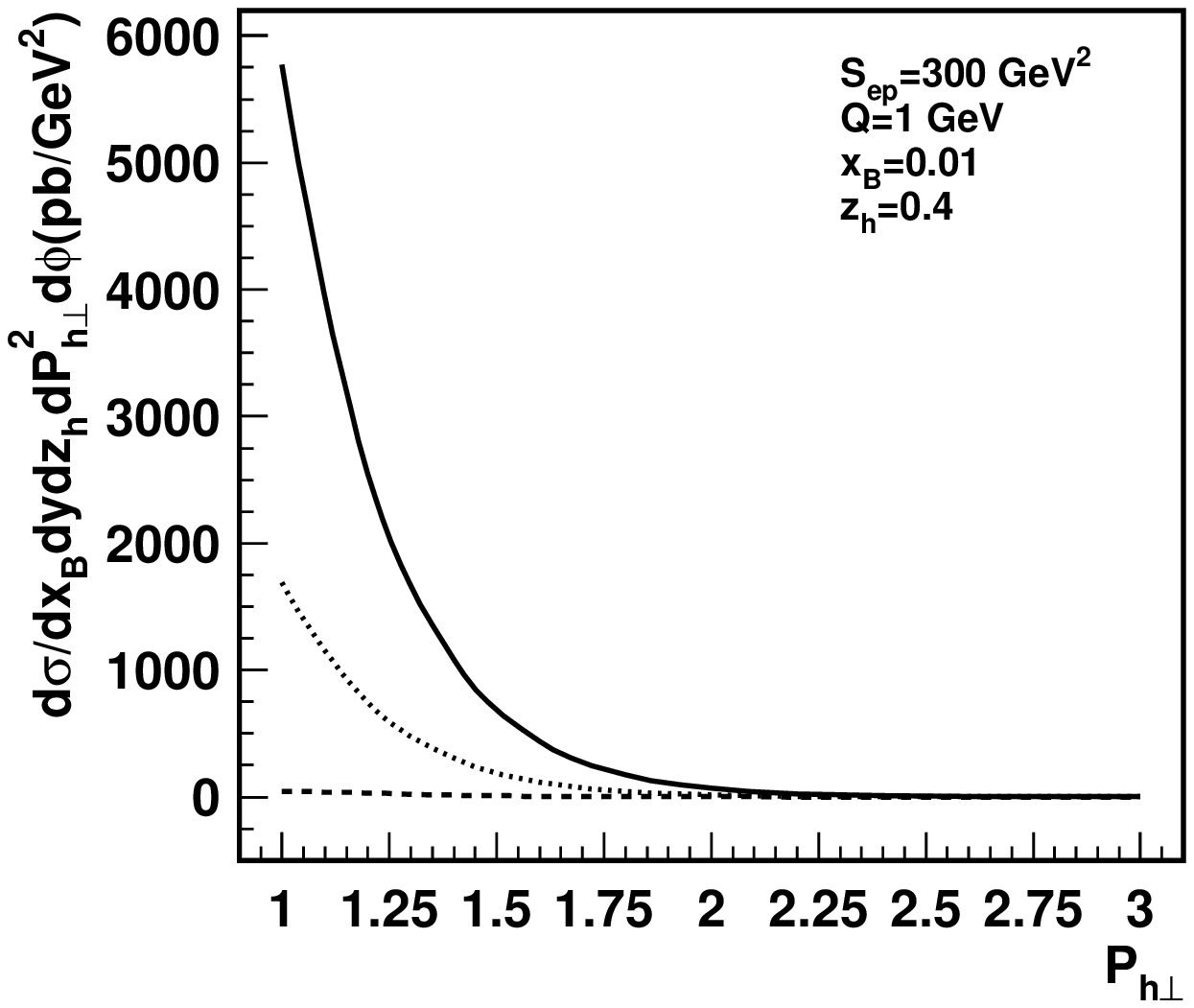,height=2.5in}
\caption{The fully differential unpolarized cross section for $D^0$ production in SIDIS for COMPASS kinematics. The curves represent: $\sigma_0^U$(solid), $\sigma_1^U$(dashed), and $\sigma_2^U$(dotted) in Eq.~(\protect\ref{angulardis}).}
\label{com-LO}
\eef

In Fig.~\ref{com-LO}, we show individual coefficients of the angular distribution, $\sigma_0^U$, $\sigma_1^U$, and $\sigma_2^U$, of the {\it fully differential} unpolarized cross section for $D^0$ production in Eq.~(\ref{angulardis}) as a function of both $z_h$ and $P_{h\perp}$ for the kinematics relevant to COMPASS experiment. It is clear that the angular dependent pieces $\sigma_1^U, \sigma_2^U\ll \sigma_0^U$, and might be too small to be significant.  Without worrying about the detection efficiency, the $D$-meson production at $P_{h\perp}\sim 1$~GeV could be measurable.  Likewise, Fig.~\ref{erhic-LO} shows the {\it fully differential} unpolarized cross section for $D^0$ production for eRHIC kinematics. With larger $Q$ and $P_{h\perp}$, the production rate is smaller but may still have enough events with a high luminosity. 
\bef
\psfig{file=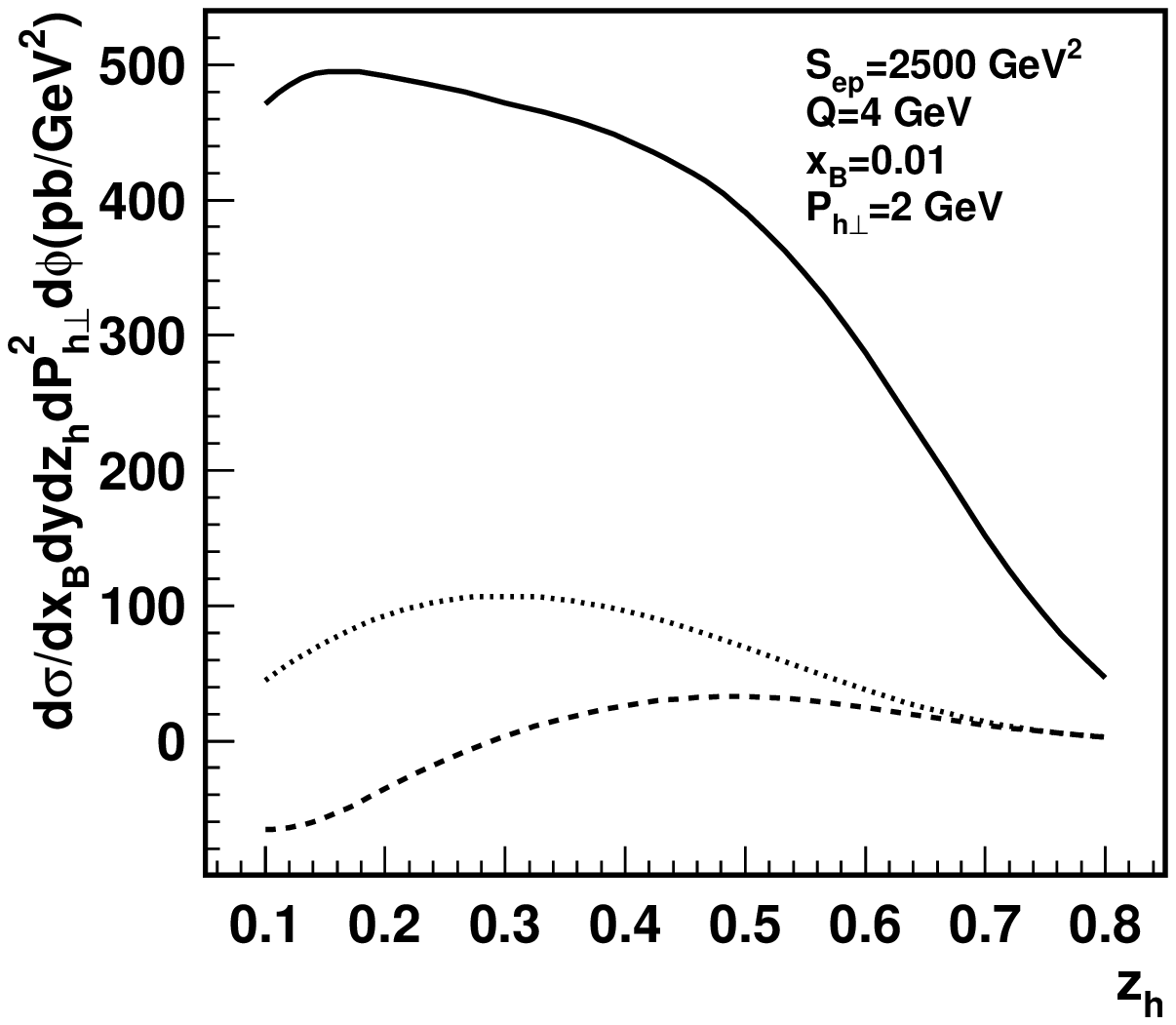,height=2.5in}
\hskip 0.2in
\psfig{file=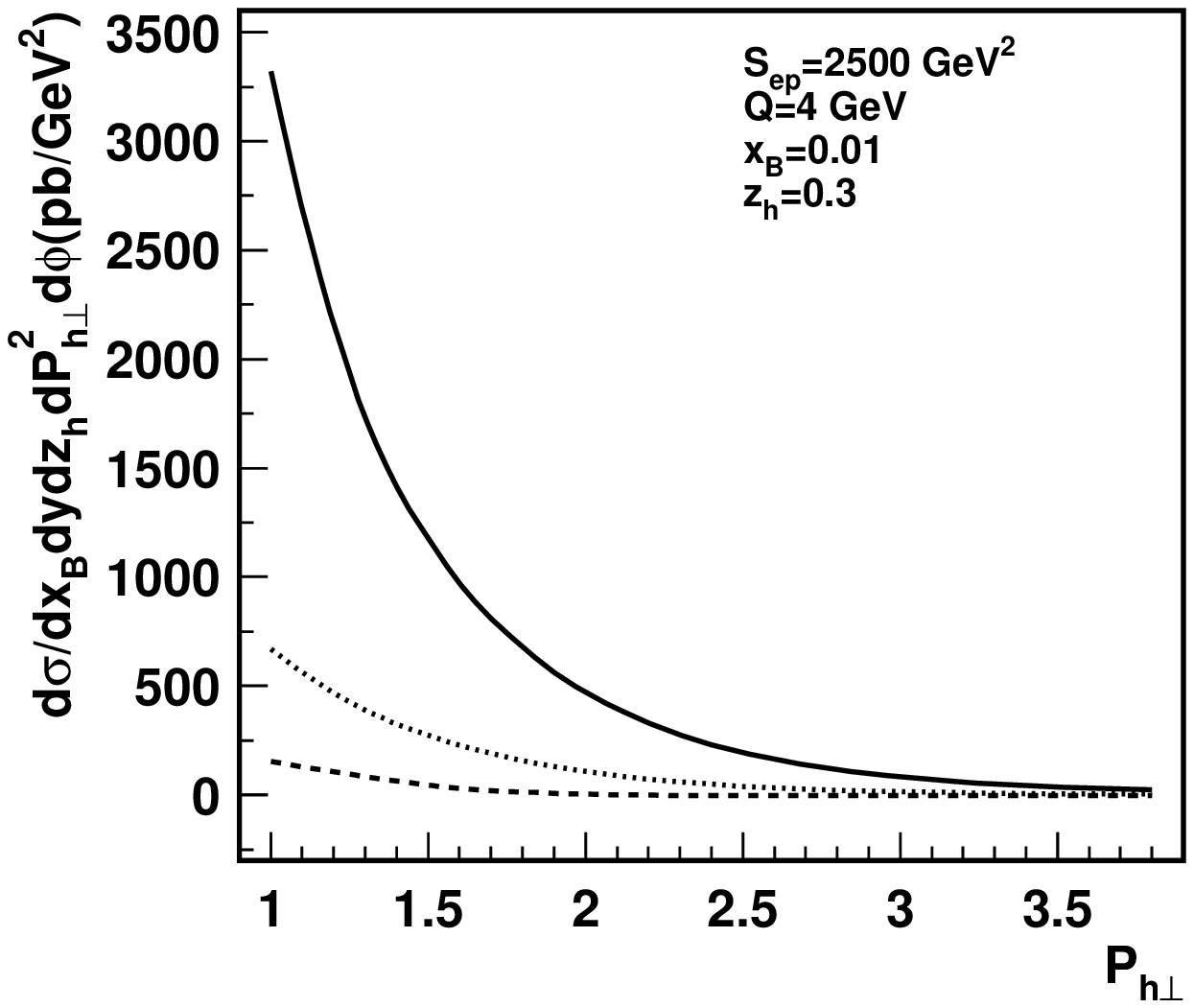,height=2.5in}
\caption{The fully differential unpolarized cross section for $D^0$ production in SIDIS at the future eRHIC. The curves represent: $\sigma_0^U$(solid), $\sigma_1^U$(dashed), and $\sigma_2^U$(dotted) in Eq.~(\protect\ref{angulardis}).}
\label{erhic-LO}
\eef

In order to obtain the numerical estimate for the SSAs of $D$-meson production, we have to model the unknown, but universal, tri-gluon correlation function $T_G(x, x)$.  Similar to the ansatz for quark-gluon correlation function $T_F(x, x)$, which was originally introduced in \cite{qiu} and found to be consistent with the latest experimental data \cite{Kouvaris:2006zy}, we model the tri-gluon correlation function $T_G(x, x)$ as
\ben
T_G(x, x)=\lambda_g\, G(x)
\een
with $G(x)$ the normal unpolarized gluon distribution function. Because of its non-perturbative nature, $T_G(x,x)$ should be extracted from the experiments and the value and the sign of $\lambda_g$ should be fixed by the data.  For the following numerical estimate, we assume that $\lambda_g$ has a positive sign and the same size as that for quark-gluon correlation function $T_F(x, x)$ \cite{qiu}, and choose $\lambda_g=0.07$~GeV.

In order to present the SSA and its angular dependence on the $\phi$, the angle between the hadron plane and the lepton plane, we define the $\phi$-integrated single spin azimuthal asymmetries as
\ben
\langle \cos(n\phi) \rangle=\frac{1}{\sin{\phi_s}}\,
\frac{\int_0^{2\pi}d\phi \cos(n\phi) 
\frac{d\Delta\sigma(s_\perp)}{dx_B dy dz_h dP_{h\perp}^2 d\phi}}
{\int_0^{2\pi}d\phi \frac{d\sigma}{dx_B dy dz_h dP_{h\perp}^2 d\phi}}\, ,
\een
which gives
\ben
\langle 1 \rangle = \frac{\Delta\sigma_0}{\sigma_0^U},
\qquad
\langle \cos\phi \rangle = \frac{\Delta\sigma_1}{2\sigma_0^U},
\qquad
\langle \cos2\phi \rangle = \frac{\Delta\sigma_2}{2\sigma_0^U}.
\label{moment}
\een

In Fig.~\ref{com-ssa} we plot the SSAs as a function of $z_h$ (left) and $P_{h\perp}$ (right) for the COMPASS kinematics. The asymmetries, $\langle 1 \rangle$, $\langle \cos\phi \rangle$, and $\langle \cos2\phi \rangle$, defined in Eq.~(\ref{moment}), are shown by the solid, dot-dashed, and dotted curves, respectively.  For a comparison between the size of the ``derivative'' and the ``non-derivative'' terms, we also show, by the dashed curves, the contribution to the SSA, $\langle 1 \rangle$, from the derivative term only.  It is clear  that the derivative term dominates over the whole kinematic region.  The asymmetries, $\langle \cos\phi \rangle$ and $\langle \cos2\phi \rangle$, are too small to be observed experimentally.  The SSA, $\langle 1 \rangle$, is of the order of $10\%$, and could be measurable at COMPASS experiment.  

Fig.~\ref{com-ssa} indicates that the SSA hits a minimum at $z_h\sim 0.5$ and increases very fast when $z_h$ becomes very large or very small.  This is because the SSA, $\langle 1 \rangle \sim 1/(1-x_{min})$, due to the derivative of $T_G(x, x)$ \cite{qiu}.  From the definition of $x_{min}$ in Eq.~(\ref{xmin}), the $z_h(1-z_h)$ has a maximum at $z_h=0.5$.  Therefore, $x_{min}$ increases, equivalently, the SSA increases when $z_h$ becomes either smaller or larger than $0.5$.  When $z_h$ is much further away from the central value 0.5, the $x_{min}$ becomes so large that the perturbatively calculated asymmetry could increase sharply, which could signal a breakdown of the twist-three approximation and a need of higher power corrections.  Nevertheless, the increase of the SSA when $z_h$ is moving away from the central value 0.5 has the same physics origin as the observed increase of the SSA as a function of increasing $x_F$ (or rapidity $y$) in the hadronic pion production \cite{SSA-fixed-tgt,SSA-rhic}, and it could be tested in the COMPASS experiment. 

Fig.~\ref{com-ssa} also indicates a monotonic increase of the SSA as a function of $P_{h\perp}$.  Although we expect the SSA to fall when $P_{h\perp}$ increases, a natural behavior of the twist-three effect in QCD collinear factorization, the enhancement from the derivative of the $T_G(x,x)$ at large $x$ wins over the suppression from large $P_{h\perp}$ due to the limited phase space at COMPASS kinematics.  As we will see below, the decrease of the SSA as the increase of $P_{h\perp}$ is clearly seen at the eRHIC kinematics.

\bef
\psfig{file=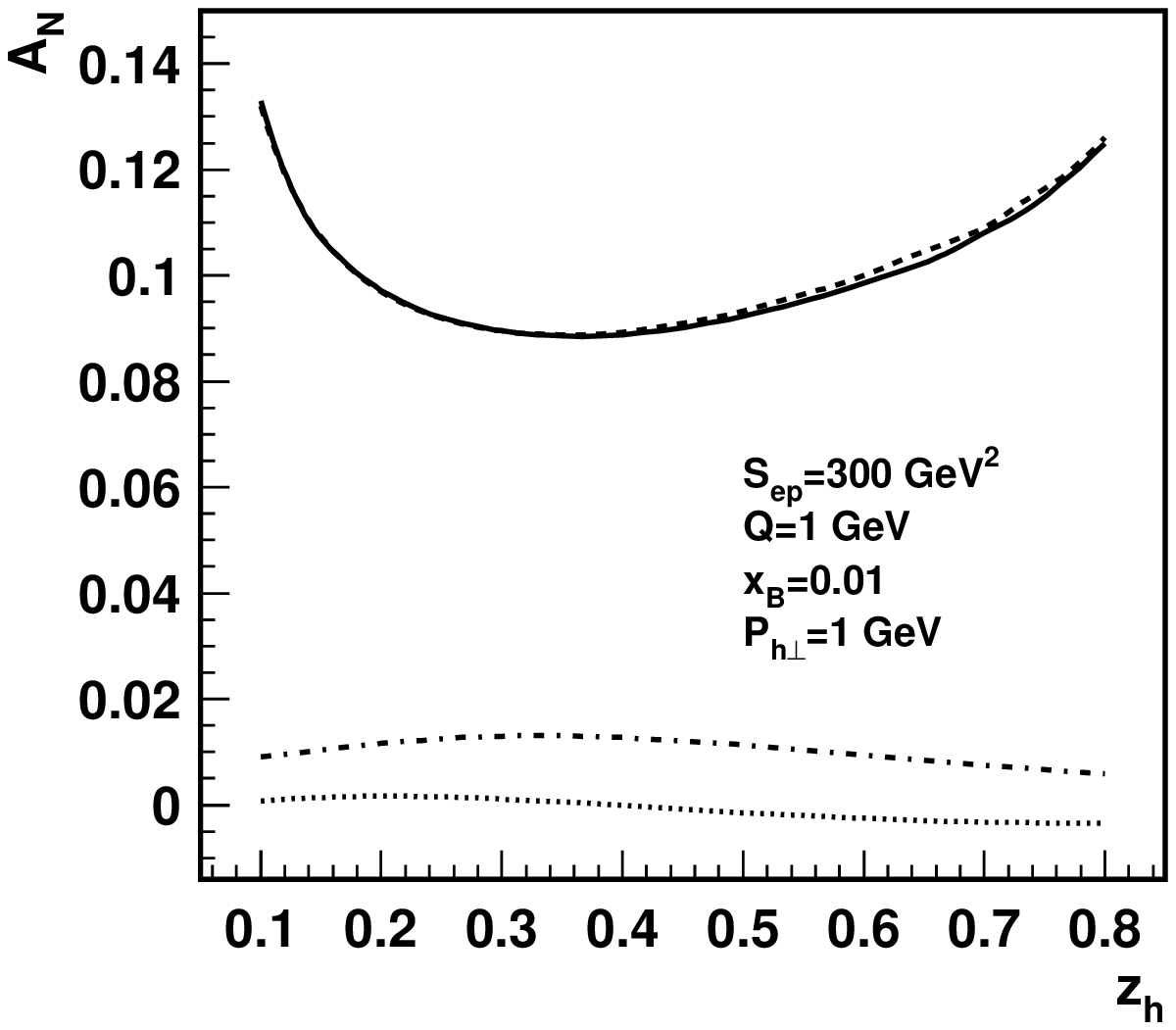,height=2.5in}
\hskip 0.2in
\psfig{file=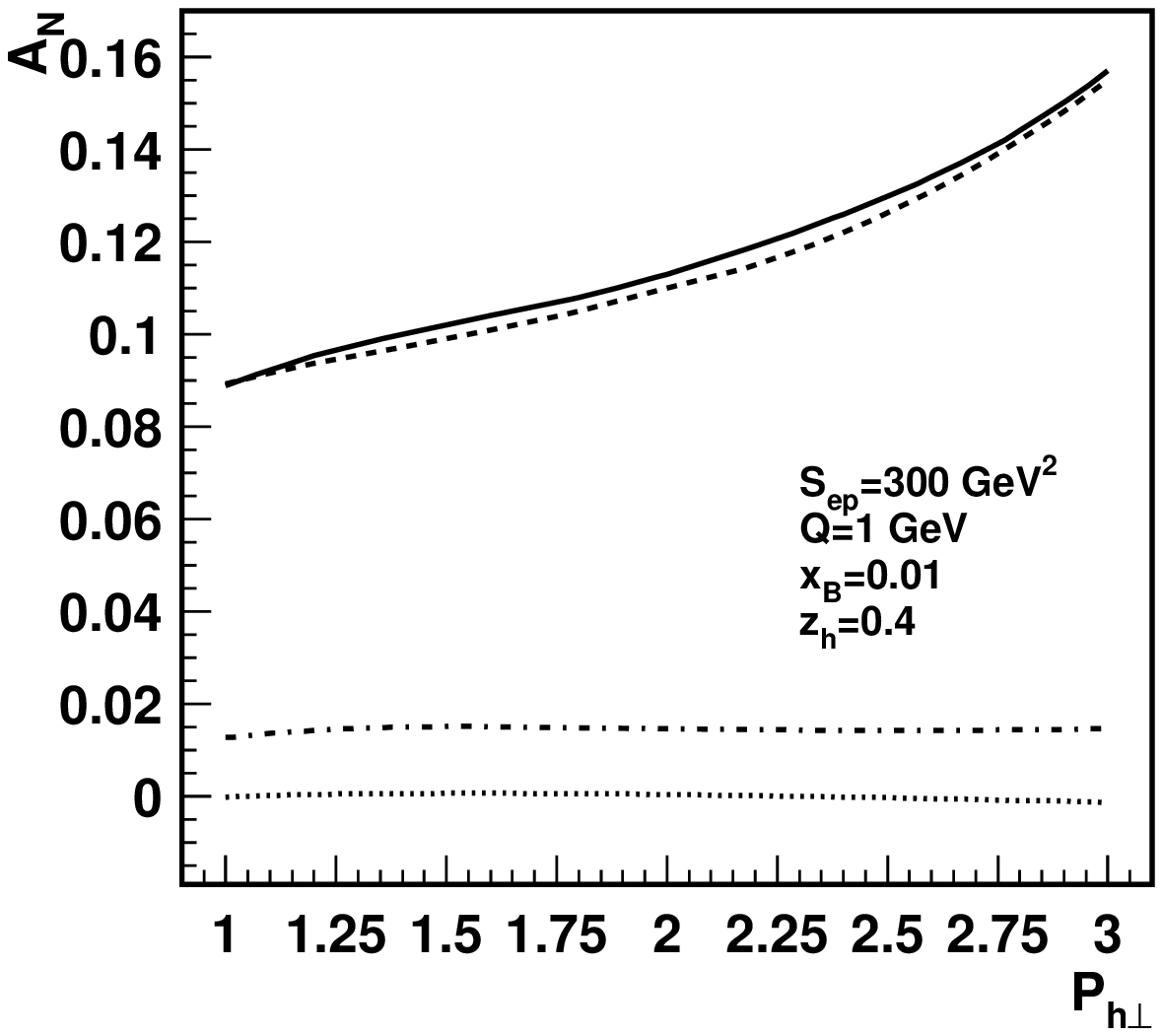,height=2.5in}
\caption{Single-transverse-spin-asymmetries defined in Eq. (\ref{moment}) for $D^0$ production in SIDIS for COMPASS kinematics. The curves are: solid-$\langle 1 \rangle$, dashed-$\langle 1 \rangle$ with derivative-term only, dot-dashed-$\langle \cos\phi \rangle$, and dotted-$\langle \cos{2\phi} \rangle$.}
\label{com-ssa}
\eef

Similarly, we plot the SSAs for $D^0$ production for the eRHIC kinematics in Fig. \ref{rhic-ssa}.  Due to the higher collision energy, the effective gluon momentum fraction $x$ that dominates the SSAs is smaller, which leads to a smaller derivative of $T_G(x, x)$ and a smaller SSAs.  Similar feature has been seen in the SSA for hadronic pion production when we compare the data from the fixed-target experiments with that from RHIC experiments.  The $5\%$ SSA for $D$-meson production at eRHIC could be significant.

The slightly different shape of the SSA as a function of $z_h$ is purely a consequence of the difference in effective range of parton momentum fraction $x$.  That is, the $z_h$-dependence of the SSA provides a good measurement of the $x$-dependence of the correlation function, $T_G(x,x)$.  On the other hand, the slow falloff of the SSA as a function $P_{h\perp}$ is natural due to the asymptotic $\lambda_g/P_{h\perp}$ behavior of the twist-3 contribution when $P_{h\perp}$ increases.  Of course, as discussed above, the $1/(1-x_{min})$ dependence of the twist-three formalism compensates some of the $1/P_{h\perp}$ falloff due to the phase space limit on parton momentum fraction $x$. 
\bef
\psfig{file=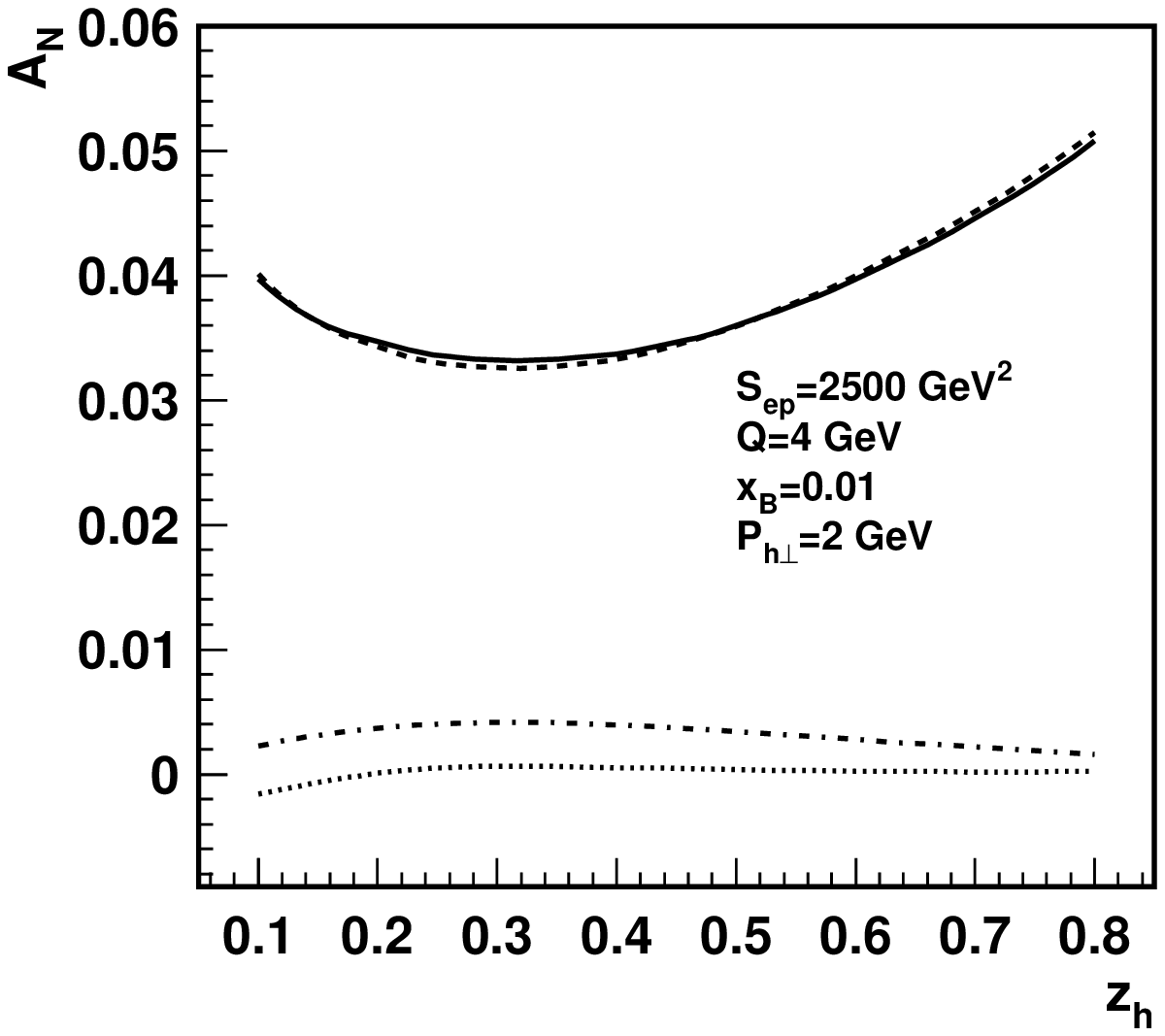,height=2.5in}
\hskip 0.2in
\psfig{file=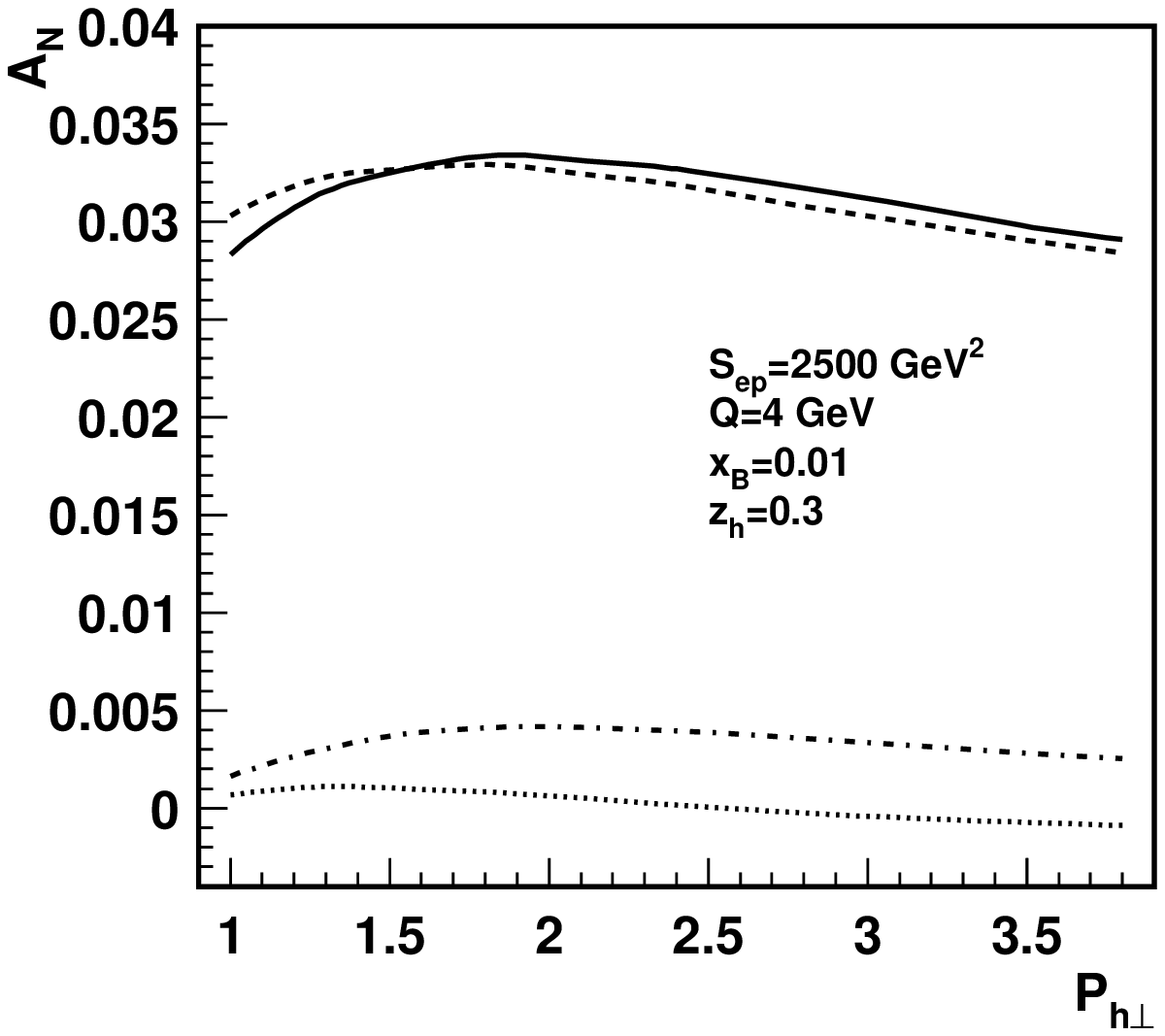,height=2.5in}
\caption{Single-transverse-spin-asymmetries defined in Eq. (\ref{moment}) for $D^0$ production in SIDIS for eRHIC kinematics. The curves are: solid-$\langle 1 \rangle$, dashed-$\langle 1 \rangle$ with derivative-term only, dot-dashed-$\langle \cos\phi \rangle$, and dotted-$\langle \cos{2\phi} \rangle$.}
\label{rhic-ssa}
\eef

\section{Conclusions}
\label{conclusion}

In summary, we have studied the single transverse-spin asymmetry for $D$-meson production in SIDIS.  In terms of QCD collinear factorization approach, we calculated both derivative and non-derivative contributions to the SSAs.  
At large enough transverse momentum, $P_{h\perp}$, the intrinsic charm contribution to the asymmetry might be neglected, and the SSA is directly proportional to the transverse-spin dependent tri-gluon correlation function, $T_G(x,x)$ (or $T_G(x,x)\pm \widetilde{T}_G(x,x)$ if we include both color structures), which has not been studied experimentally.  We pointed out
that by comparing the SSAs for producing $D$ and $\bar{D}$ mesons in SIDIS, we could gain valuable information on both tri-gluon correlation functions. With a simple model for the $T_G(x,x)$, we presented our estimates of the SSAs for the kinematics relevant for both COMPASS and future eRHIC experiments.  From the inclusive production rate for the $D$-meson production and the estimated size of the SSAs, we argue that the SSAs of $D$-meson production in SIDIS could be a direct and clean probe of the unknown tri-gluon correlation function, $T_G(x,x)$, which provides the important information on the spin-depenence of gluon's transverse motion inside a polarized hadron.  

However, we stress that the SSAs shown in all figures are directly proportional to the value and the sign of the $\lambda_g$ and our model for the twist-three tri-gluon correlation function, $T_G(x,x)$.  A different $x$-dependence from that of $G(x)$ could lead to a different derivative of $T_G(x,x)$ and a different prediction for the SSA. The actual sign and size of the SSA, and the function, $T_G(x,x)$, should be determined by the experimental measurements, just like the PDFs. However, our calculation does predict the short-distance dynamics and the kinematic dependence of the SSAs, such as the increase of the SSA when $z_h$ moves away from the central value $0.5$.

Finally, we emphasize that the QCD collinear factorization approach to the SSAs allows us to calculate the SSAs of open charm production or other particle production in hadronic collisions.  With the experimental extraction of the tri-gluon correlation function, $T_G(x,x)$, as well as $\widetilde{T}_G(x,x)$, and the existing and new knowledge of $T_F(x,x)$, we will be able to explore non-perturbative physics, in particular, the multi-parton quantum correlations, beyond what have learned from the parton distribution functions.

\section*{Acknowledgments}
We thank Werner Vogelsang and Feng Yuan for helpful discussion and thank G. Kramer for providing us with their fortran code for the $D$-meson fragmentation functions. This work was supported in part by the U. S. Department of Energy under Grant No.~DE-FG02-87ER40371. 



\begin{thebibliography}{99}

\bibitem{SSA-fixed-tgt}
D.~L.~Adams {\it et al.}  [E581 and E704 Collaborations],
  Phys.\ Lett.\ B {\bf 261}, 201 (1991);
 D.~L.~Adams {\it et al.}  [FNAL-E704 Collaboration],
  Phys.\ Lett.\ B {\bf 264}, 462 (1991);
K.~Krueger {\it et al.}, Phys.\ Lett.\ B {\bf 459}, 412 (1999).

\bibitem{SSA-dis}
A.~Bravar  [Spin Muon Collaboration],
Nucl.\ Phys.\ A {\bf 666}, 314 (2000);
A.~Airapetian {\it et al.}  [HERMES Collaboration],
Phys.\ Rev.\ Lett.\  {\bf 84}, 4047 (2000);
Phys.\ Rev.\ D {\bf 64}, 097101 (2001);
  Phys.\ Rev.\ Lett.\  {\bf 94}, 012002 (2005)
  [arXiv:hep-ex/0408013];
  V.~Y.~Alexakhin {\it et al.}  [COMPASS Collaboration],
  Phys.\ Rev.\ Lett.\  {\bf 94}, 202002 (2005)
  [arXiv:hep-ex/0503002];
  E.~S.~Ageev {\it et al.}  [COMPASS Collaboration],
  Nucl.\ Phys.\  B {\bf 765}, 31 (2007)
  [arXiv:hep-ex/0610068];
  M.~Alekseev {\it et al.}  [COMPASS Collaboration],
  arXiv:0802.2160 [hep-ex];
  H.~Avakian, P.~E.~Bosted, V.~Burkert and L.~Elouadrhiri  [CLAS
                  Collaboration],
  AIP Conf.\ Proc.\  {\bf 792}, 945 (2005)
  [arXiv:nucl-ex/0509032].

\bibitem{SSA-rhic}
  J.~Adams {\it et al.}  [STAR Collaboration],
  Phys.\ Rev.\ Lett.\  {\bf 92}, 171801 (2004)
  [arXiv:hep-ex/0310058];
B.~I.~Abelev {\it et al.}  [STAR Collaboration],
  Phys.\ Rev.\ Lett.\  {\bf 99}, 142003 (2007)
  [arXiv:0705.4629 [hep-ex]];
  arXiv:0801.2990 [hep-ex];
  S.~S.~Adler {\it et al.}  [PHENIX Collaboration],
  Phys.\ Rev.\ Lett.\  {\bf 95}, 202001 (2005)
  [arXiv:hep-ex/0507073];
  I.~Arsene {\it et al.}  [BRAHMS Collaboration],
  arXiv:0801.1078 [nucl-ex].
  
  
\bibitem{SSA-review}
for reviews, see:
M.~Anselmino, A.~Efremov and E.~Leader,
Phys.\ Rept.\  {\bf 261}, 1 (1995) [Erratum-ibid.\  {\bf 281}, 399
(1997)];
  Z.~t.~Liang and C.~Boros,
  Int.\ J.\ Mod.\ Phys.\ A {\bf 15}, 927 (2000);
V.~Barone, A.~Drago and P.~G.~Ratcliffe,
Phys.\ Rept.\  {\bf 359}, 1 (2002);
  U.~D'Alesio and F.~Murgia,
  arXiv:0712.4328 [hep-ph].


\bibitem{Siv90}
D.~W.~Sivers,
Phys.\ Rev.\ D {\bf 41}, 83 (1990);
Phys.\ Rev.\ D {\bf 43}, 261 (1991).

\bibitem{Efremov}
  A.~V.~Efremov and O.~V.~Teryaev,
  Sov.\ J.\ Nucl.\ Phys.\  {\bf 36}, 140 (1982)
  [Yad.\ Fiz.\  {\bf 36}, 242 (1982)];
  A.~V.~Efremov and O.~V.~Teryaev,
  Phys.\ Lett.\ B {\bf 150}, 383 (1985).

\bibitem{qiu}
J.~W.~Qiu and G.~Sterman,
Phys.\ Rev.\ Lett.\  {\bf 67}, 2264 (1991);
  Nucl.\ Phys.\ B {\bf 378}, 52 (1992);
J.~W.~Qiu and G.~Sterman,
Phys.\ Rev.\ D {\bf 59}, 014004 (1999).

\bibitem{Kanazawa:2000hz}
  Y.~Kanazawa and Y.~Koike,
  Phys.\ Lett.\ B {\bf 478}, 121 (2000);
  Phys.\ Rev.\ D {\bf 64}, 034019 (2001).

\bibitem{Vogelsang:2005cs}
  W.~Vogelsang and F.~Yuan,
  Phys.\ Rev.\  D {\bf 72}, 054028 (2005)
  [arXiv:hep-ph/0507266].

\bibitem{siverscompare}
  M.~Anselmino {\it et al.},
  arXiv:hep-ph/0511017, and references therein.

\bibitem{Ans94}
M.~Anselmino, M.~Boglione and F.~Murgia,
Phys.\ Lett.\ B {\bf 362}, 164 (1995);
M.~Anselmino and F.~Murgia,
Phys.\ Lett.\ B {\bf 442}, 470 (1998);
U.~D'Alesio and F.~Murgia,
Phys.\ Rev.\ D {\bf 70}, 074009 (2004);
M.~Anselmino, M.~Boglione, U.~D'Alesio, E.~Leader, S.~Melis and F.~Murgia,
Phys.\ Rev.\ D {\bf 73}, 014020 (2006).

\bibitem{MulTanBoe}
  P.~J.~Mulders and R.~D.~Tangerman,
  Nucl.\ Phys.\ B {\bf 461}, 197 (1996)
  [Erratum-ibid.\ B {\bf 484}, 538 (1997)];
  D.~Boer and P.~J.~Mulders,
  Phys.\ Rev.\ D {\bf 57}, 5780 (1998).

\bibitem{Boer:2003tx}
  D.~Boer and W.~Vogelsang,
  Phys.\ Rev.\  D {\bf 69}, 094025 (2004)
  [arXiv:hep-ph/0312320].

\bibitem{Qiu:2007ar}
  J.~W.~Qiu, W.~Vogelsang and F.~Yuan,
  Phys.\ Lett.\  B {\bf 650}, 373 (2007)
  [arXiv:0704.1153 [hep-ph]];
  Phys.\ Rev.\  D {\bf 76}, 074029 (2007)
  [arXiv:0706.1196 [hep-ph]].

\bibitem{Boer:2003cm}
  D.~Boer, P.~J.~Mulders and F.~Pijlman,
  Nucl.\ Phys.\  B {\bf 667}, 201 (2003)
  [arXiv:hep-ph/0303034];
  J.~P.~Ma and Q.~Wang,
  Eur.\ Phys.\ J.\  C {\bf 37}, 293 (2004)
  [arXiv:hep-ph/0310245];
  A.~Bacchetta,
  arXiv:hep-ph/0511085.

\bibitem{UnifySSA}
  X.~Ji, J.~W.~Qiu, W.~Vogelsang and F.~Yuan,
  Phys.\ Rev.\ Lett.\  {\bf 97}, 082002 (2006)
  [arXiv:hep-ph/0602239],
  Phys.\ Rev.\  D {\bf 73}, 094017 (2006)
  [arXiv:hep-ph/0604023],
  Phys.\ Lett.\  B {\bf 638}, 178 (2006)
  [arXiv:hep-ph/0604128];
  Y.~Koike, W.~Vogelsang and F.~Yuan,
  Phys.\ Lett.\  B {\bf 659}, 878 (2008)
  [arXiv:0711.0636 [hep-ph]].

\bibitem{Kouvaris:2006zy}
  C.~Kouvaris, J.~W.~Qiu, W.~Vogelsang and F.~Yuan,
  Phys.\ Rev.\  D {\bf 74}, 114013 (2006)
  [arXiv:hep-ph/0609238].

\bibitem{Anselmino:2004nk}
  M.~Anselmino, M.~Boglione, U.~D'Alesio, E.~Leader and F.~Murgia,
  Phys.\ Rev.\  D {\bf 70}, 074025 (2004)
  [arXiv:hep-ph/0407100];
  M.~Anselmino, U.~D'Alesio, S.~Melis and F.~Murgia,
  Phys.\ Rev.\  D {\bf 74}, 094011 (2006)
  [arXiv:hep-ph/0608211].

\bibitem{Ji:1992eu}
  X.~D.~Ji,
  Phys.\ Lett.\  B {\bf 289}, 137 (1992).

\bibitem{Brodsky:ic}
  S.~J.~Brodsky, P.~Hoyer, C.~Peterson and N.~Sakai,
  Phys.\ Lett.\  B {\bf 93}, 451 (1980);
  S.~J.~Brodsky, C.~Peterson and N.~Sakai,
  Phys.\ Rev.\  D {\bf 23}, 2745 (1981).

\bibitem{compass}
  M.~Alekseev {\it et al.}  [COMPASS Collaboration],
  arXiv:0802.3023 [hep-ex].

\bibitem{Deshpande:2005wd}
  A.~Deshpande, R.~Milner, R.~Venugopalan and W.~Vogelsang,
  Ann.\ Rev.\ Nucl.\ Part.\ Sci.\  {\bf 55}, 165 (2005)
  [arXiv:hep-ph/0506148].

\bibitem{Koike:proof}
  Y.~Koike and K.~Tanaka,
  Phys.\ Lett.\  B {\bf 646}, 232 (2007)
  [arXiv:hep-ph/0612117];
  Phys.\ Rev.\  D {\bf 76}, 011502 (2007)
  [arXiv:hep-ph/0703169].

\bibitem{Yuan:2008it}
  F.~Yuan and J.~Zhou,
  arXiv:0806.1932 [hep-ph].

\bibitem{Vitev:2006bi}
  I.~Vitev, J.~T.~Goldman, M.~B.~Johnson and J.~W.~Qiu,
  Phys.\ Rev.\  D {\bf 74}, 054010 (2006)
  [arXiv:hep-ph/0605200].

\bibitem{Kang:hadron}
Z.~B.~Kang, in preparation.

\bibitem{sidis}
  R.~b.~Meng, F.~I.~Olness and D.~E.~Soper,
  Nucl.\ Phys.\ B {\bf 371}, 79 (1992);
  Y.~Koike and J.~Nagashima,
  Nucl.\ Phys.\ B {\bf 660}, 269 (2003).

\bibitem{Mendez:1978zx}
  A.~Mendez,
  Nucl.\ Phys.\  B {\bf 145}, 199 (1978).

\bibitem{Koike}
  H.~Eguchi, Y.~Koike and K.~Tanaka,
  Nucl.\ Phys.\  B {\bf 763}, 198 (2007)
  [arXiv:hep-ph/0610314].

\bibitem{Qiu:1990cu}
  J.~W.~Qiu and G.~Sterman,
  AIP Conf.\ Proc.\  {\bf 223}, 249 (1991).

\bibitem{Yuan:private}
F.~Yuan, private communication.

\bibitem{Pumplin:2002vw}
  J.~Pumplin, D.~R.~Stump, J.~Huston, H.~L.~Lai, P.~Nadolsky and W.~K.~Tung,
  JHEP {\bf 0207}, 012 (2002)
  [arXiv:hep-ph/0201195].

\bibitem{Kneesch:2007ey}
  T.~Kneesch, B.~A.~Kniehl, G.~Kramer and I.~Schienbein,
  arXiv:0712.0481 [hep-ph].

 
\end{thebibliography}
\end{document}